# Perturbation Biology: inferring signaling networks in cellular systems

Evan J. Molinelli*[2,3], Anil Korkut*[2], Wei-Qing Wang*[2], Martin L. Miller[2], Nicholas P. Gauthier[2], Xiaohong Jing[2], Poorvi Kaushik[2,3], Qin He[2], Gordon Mills[4], David B. Solit[5,6], Christine A. Pratilas[5,7], Martin Weigt[1], Alfredo Braunstein[1], Andrea Pagnani[1], Riccardo Zecchina[1], Chris Sander[2,4]

[1]Politecnico di Torino and Human Genetics Foundation, HuGeF, Torino, Italy
[2]Computational Biology Program, Memorial Sloan-Kettering Cancer Center, New York, USA
[3]Tri-Institutional Program for Computational Biology and Medicine, Weill Cornell Medical College, New York, USA
[4]Department of Systems Biology, The University of Texas MD Anderson Cancer Center, Houston, TX, USA
[5]Program in Molecular Pharmacology, Memorial Sloan-Kettering Cancer Center, New York, NY, USA
[6]Human Oncology and Pathogenesis Program, Memorial Sloan-Kettering Cancer Center, New York, NY, USA
[7]Department of Pediatrics, Memorial Sloan-Kettering Cancer Center, New York, NY, USA

* Joint first authors

[4]Email to perturbation-biology@cbio.mskcc.org reaches the principal authors

## Abstract

We present a new experimental-computational technology of inferring network models that predict the response of cells to perturbations and that may be useful in the design of combinatorial therapy against cancer. The experiments are systematic series of perturbations of cancer cell lines by targeted drugs, singly or in combination. The response to perturbation is measured in terms of levels of proteins and phospho-proteins and of cellular phenotype such as viability. Computational network models are derived de novo, i.e., without prior knowledge of signaling pathways, and are based on simple non-linear differential equations. The prohibitively large solution space of all possible network models is explored efficiently using a probabilistic algorithm, belief propagation, which is three orders of magnitude more efficient than Monte Carlo methods. Explicit executable models are derived for a set of perturbation experiments in Skmel-133 melanoma cell lines, which are resistant to the therapeutically important inhibition of Raf kinase. The resulting network models reproduce and extend known pathway biology. They can be applied to discover new molecular interactions and to predict the effect of novel drug perturbations, one of which is verified experimentally. The technology is suitable for application to larger systems in diverse areas of molecular biology.

**Author Summary**

Drugs that target specific behaviors of signaling proteins are a promising new strategy for treating cancer. One of the many obstacles facing optimal drug design is the incomplete understanding of the coordinated interactions between the signaling proteins at a quantitatively predictive level. *De Novo* model inference refers to the algorithmic construction of mathematical models from experimental data without dependence on prior knowledge, in an effort to model these signaling pathways in a specific genetic context. *De Novo* inference is difficult because of the prohibitively large number of possible arrangements of interactions. Our lab has adapted a method called Belief Propagation that calculates probabilistically the most likely interactions, from which we can explore a large number of highly probable models. In this paper, we test this method on artificial data and then apply it to model signaling pathways in a BRAF-mutant melanoma cancer cell line. Our results show agreement with known biology, predict novel interactions, and predict efficacious drug targets, specific to the experimental cell line. We believe that this methodology has the potential, with sufficient data, to model upwards of 100 proteins, a scale so far unreachable with current technology.

# Introduction

**Signaling in cancer cells**. Abnormal bio-molecular information flow as a result of genetic or epigenetic alterations may lead to tumorigenic transformation and malignancy and is classically modeled as changes in signaling pathways [1]. Targeted anti-cancer drugs, which bind and inhibit specific components of aberrant signaling pathways, are a promising alternative to conventional chemotherapy, with recent successes in melanoma (RAF inhibitor) [2] and prostate cancer (AR inhibitor) [3,4] following in the footsteps of the pioneering bcr/abl inhibitor Imatinib [5] and EGFR inhibitors Gefitinib and Erlotinib [6,7,8]. Combinations of targeted anticancer drugs hold considerable promise because of emergence of resistance to initially successful single agents and the highly robust nature of the signaling pathways with multiple feedback mechanisms [9].

**Data-driven models of cell biology.** High throughput measurements on response profiles of living cells to multiple perturbations such as drug combinations provide a rich set of information to construct quantitative cell biology models. In this paper, we construct context specific de novo mathematical models of signaling pathways through use of systematic paired perturbation experiments and network inference algorithms. Such network models will serve to gain insight into mechanistic details of signaling pathways, predict the response of cellular systems to multiple perturbations beyond the conditions in which models are generated and will be useful for design of perturbations that achieve a desired response.

**State of the art in network inference in cell biology.** Previous mathematical models of molecular signaling in cells have been effective in modeling pathways and enhancing drug discovery [10,11,12,13,14,15,16,17,18,19]. Techniques for network modeling of signaling pathways span a spectrum of approaches at different levels of complexity. Modeling techniques in terms of detailed chemical kinetics and spatiotemporal models [17,20,21] can provide mechanistic explanations of observed behavior, but are often incompletely parameterized, needing tens or hundreds of kinetic parameters for medium-sized systems. Moreover, such models may not be valid in biological contexts that differ substantially from dilute solution chemistry. On the other end of the spectrum, pattern matching or machine learning models such as neural networks and correlation-based models such as maximum entropy [22] can accurately provide purely data-driven models of signaling. However, such methods have limited power to explain mechanistic details and, in most cases, are insufficient for quantitative predictions of system behavior in conditions beyond those from which the models are derived.

**Data-driven and context-specific predictive models**. Here, we take a unique modeling approach to construct context specific, de novo and predictive network models of signaling pathways from drug perturbation data (Figure 1). Here, de novo means that network inference is done without depending on known molecular interactions, e.g., extracted from literature or pathway databases that are not curated to account for biological context. This approach also emphasizes context specificity since it relies on rich experimental data from a single biological context as its training set. The models are constructed through parameterization of a simple model equation, which has been used in other network modeling approaches [13,23,24,25,26,27]. The ansatz of the model equation yields parameters that are both mechanistically descriptive of system interactions (Supplemental Figure S15) and computationally predictive of cell-type specific response to new drug perturbations and perturbation combinations. We expect that the union of these advantages, will empower the community to identify unique drug targets and combinations that are particularly efficacious to the given context.

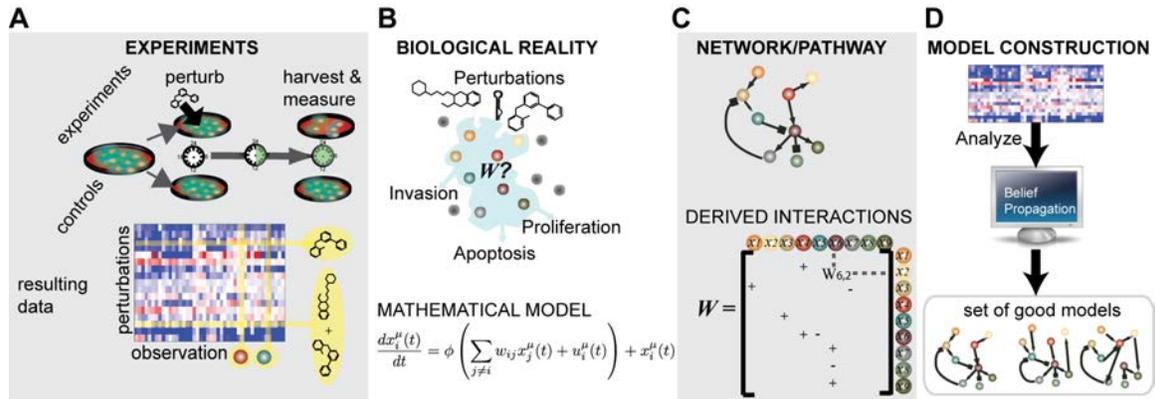

**Figure 1: Perturbation cell biology**. Perturbing cancer cells with targeted drugs singly and in pairs (A) reveals context-specific response to therapies and illuminates protein interactions. We construct dynamic mathematical models of the cells' response to drugs that have both quantitative parameters (B) and a qualitative network interpretation (C). We use an inference algorithm called Belief Propagation (BP) to construct a set of good, i.e., predictive models (D).

**Network modeling de novo or with prior information**. *De novo* construction of signaling network models at scales relevant to problems related to complex biological phenomena such as cancer has long been a challenge in system biology. Thus, quantitative models of protein signaling pathways are typically constructed on the basis of existing prior knowledge from literature searches [16,17] and interaction databases [19,28,29]. However, different cancer contexts have unique genetic and proteomic alterations to normal protein signaling. Moreover, signaling pathways have a highly non-linear nature with negative feedback loops, non-overlapping pathways and cross-talk etc. For example, distinct mutations in effector proteins of the PI3K pathway confer oncogenicity in unique ways leading to highly context dependent functional consequences across different cancer types, subtypes and patients [30]. The method we introduce in this document is capable of inferring parameterized network models with or without prior knowledge. We test our method without prior knowledge. An advantage of *de novo* inference is the liberation from prior knowledge interactions that may be incorrect in a given biological context.

**Model inference from perturbation data for larger systems is hard**. The largest obstacle to *de novo* model construction is the combinatorial explosion in the number of possible networks [31]. Model inference problems of this type are NP-hard [32]. The number of possible configurations for a model with N nodes, and K possible values for each parameter grows super-exponentially as $K^{N \times N}$. We have previously described a method for *de novo* construction of dynamic nonlinear network models from perturbation data [33]. This network construction method is based on the combined use of a Monte Carlo stochastic search algorithm, that is used to find the network topology and an efficient gradient descent algorithm [34] for quantitative parameter determination. However, without algorithmic improvements, such Monte Carlo based methods are limited to modeling fewer than approximately 15-20 biological entities [33]. Increasing the scale of network (or pathway) models of cellular signaling processes to levels sufficient to describe complex biological problems in quantitative detail is therefore extremely challenging and has been approach with a diversity of methods [35].

**A statistical physics approach can handle the complexity for larger systems**. An ingenious, two-step approach to deal with larger systems is based on first calculating probability distributions for each possible interaction in the model and then computing distinct solutions by sampling these probability distributions. For this purpose, we devised a probability model of network configurations inspired from statistical physics principles. Using a set of approximations to simplify the probability model, we adapted an iterative algorithm called belief propagation (BP) that involves local optimization updates of inferred model parameters to eventually reach a globally optimized network model solution [36,37,38]. BP has been applied to various complex inference problems, some of them NP-Hard such as K-SAT [38] and graph coloring [39]. BP has garnered some attention in biological network inference [40,41] and parameter estimation [42] . Here, we have tailored the BP algorithm to large-scale perturbation data that is capable of increasing the scope of the models to hundreds of nodes. The result of belief propagation is a set of probability distributions for each model parameter, often referred to as marginals. Each marginal describes the inferred distribution of a particular parameter across a range of 'good' solutions. Individual models are created via the belief propagation guided decimation algorithm [43]. Consequently, the time-complexity of the problem is strongly reduced, the prohibitive computational demands of combinatorial complexity are circumvented and, although the method provides only an approximate solution, one obtains useful, non-trivial results.

**In practice: from systematic perturbation to response profiling to network model inference.** Our algorithmic network pharmacology approach involves four major steps: (i) perturbation experiments with combinations of targeted compounds, (ii) high-throughput proteomic measurements, e.g., with reverse phase protein array (RPPA) technology or mass spectrometry, and quantitative measurements of phenotypic features such as cell viability, (iii) inference of quantitative network models of signaling that link changes in proteomic profiles to changes in phenotypic profiles in response to single or multiple perturbations in specific biological contexts, and (iv) use of the network models to predict cellular and molecular responses to diverse perturbations, beyond the conditions on which the network models are derived.

**Experimental and computational technology for network model inference and application to drug effects on melanoma cell lines.** Here, we have adapted a probabilistic inference strategy to construct quantitative network models of signaling pathways from systematic perturbation experiments. First, we evaluated the speed and accuracy of BP on biologically inspired toy models. The inference of toy models, which

mimic biological pathways, reveals that BP offers a significant improvement in computational efficiency compared to traditional Monte Carlo simulations without a sacrifice in accuracy. Next, we have constructed network models of signaling in a RAF inhibitor resistant melanoma cell line (SKMEL-133), which has the BRAFV600E mutation [44]. Finally, we have successfully predicted the proteomic and phenotypic response profiles of melanoma cells to multiple combinations of targeted drugs using numerical simulations with *in silico* perturbations. With the introduction of many novel targeted drugs and patient specific genomic profiling, the described network pharmacology approach aims provide an effective tool to develop individualized, combinatorial therapeutic approaches against multiple cancer forms.

# RESULTS

## Theory

### *Mathematical form of the network model*

A key decision in modeling a biological cellular system is the choice of variables and the mathematical framework for representing system dynamics. Here, we work with a fairly simple but powerful *ansatz* or framework, in which the time behavior of the cellular system { $x_i(t)$ } in a set of perturbation conditions { $u_i^\mu$ } is modeled as a series of coupled non-linear differential equations (Equation 1) [33].

*Equation 1: Non-linear network model for the time behavior of the cellular system*

$$\frac{dx_i^\mu(t)}{dt} = \varepsilon_i \varphi\left(\sum_{j \neq i}^N w_{ij} x_j^\mu(t) + u_i^\mu\right) - \alpha_i x_i^\mu(t)$$

$$\varphi(z) = tanh(z)$$

Here, { $x_i(t)$ } denotes the set of temporal trajectories for all $x_i$, $i$ is an index over the network nodes, and $\mu$ is an index over the different perturbation conditions. Perturbations consist of targeted drugs applied alone or in pairs, each of which inhibits the kinase activity or function of specific proteins. In most cases the kinase activity is not directly measureable, and thus kinase inhibitors are incorporated into the model as so-called activity nodes, which couple the force of external perturbation throughout the whole system (See Methods section).

Each network node is modeled as a variable corresponding to a real number value. This value quantifies the relative concentrations of biological molecules (proteins, phospho-proteins or RNA molecules) or phenotypic attributes such as growth. In this framework the value of $x_i^\mu$ is the log2 ratio of the quantity of the $i^{th}$ entity in perturbation condition $\mu$ relative to the unperturbed condition. Consequently, model variables can take both negative and positive values denoting decreased or increased quantities of the corresponding biological entity. In the absence of absolute concentrations, we chose to normalize protein measurements against unperturbed levels so as to focus on signaling differences due directly to perturbation.

The rate of change for any variable is predominantly driven by the additive linear combination of upstream nodes $\{x_j\}$ weighted by their respective parameters $\{w_{ij}\}$. Non-zero values of $w_{ij}$ denote interactions in the network model. We incorporate nonlinearity with a sigmoidal function $\phi(z)$ that limits both the maximum positive and negative rate of change [24]. The $\alpha_i$ parameter models the rate of restoration at which a model variable would return to its initial value before perturbation, absent interactions. This is analogous to the degradation rate in models of positively valued protein concentrations. The $\varepsilon_i$ parameter tunes the saturation levels for each model variable. The parameters $\varepsilon_i$ and $\alpha_i$ are not inferred with BP, and thus to reduce the mathematical notation are assumed to be 1 for the remainder of this section.

After perturbation from some initial state at $t=t_0$, the system variables $x_i^\mu(t)$ in each condition $\mu$ evolve in time and, in general, reach a steady state at time $t=t_s$. The variables $u_i$ model the external influence of a perturbing agent (such as a drug) on model node $x_i$. In principle, $u_i$ is time-dependent, but not in the current implementation. A perturbing agent acting on a set of target nodes $\{x_i\}$ is represented as a vector with non-zero values for $\{u_i\}$. Similarly, a combination of drugs is modeled as the simple sum of such perturbation vectors.

The network models are parameterized by the interaction matrix $W = \{w_{ij}\}$, where $w_{ij}$ represents a directed interaction between nodes, quantifying the influence of $x_j$ on the rate of change of $x_i$. In chemical kinetics, the $w_{ij}$ is analogous to rate constants in units of inverse time, although no explicit rates are derived here. Equation 1 describes the dynamic behavior of the system, given a constant interaction matrix $W$. These apparently simple models can represent biologically realistic regulatory motifs, such as serial and parallel pathway connectivity, positive and negative feedback loops and feed-forward control.

Our models are dynamic in the sense that they produce simulated trajectories that settle at a final steady state. However, in this work, the parameters are inferred based solely on steady state data. Consequently, the trajectories of the model simulations are unconstrained to experimental measurements.

### *The problem of model inference*
The problem of deducing useful models of a (biological) system is called 'model inference'. The objective of model inference, given a mathematical framework like that described above, is to find a set of parameters such that the model equations best reproduce a training set of experimental data and have predictive power beyond the training set. In the present modeling framework, we aim to find numerical values for the $N^2$ parameters in the interaction matrix $W$, such that descriptive and predictive power of the model is optimized. Genuine predictive power, rather than just descriptive power, requires both low *error* and low *complexity* of the model, combined as low *cost*. More precisely, we quantify the cost of $W$ by an objective cost function $C(W)$ that penalizes (a) discrepancies between predicted ($x_i^\mu$) and experimental ($x_i^{\mu*}$) values of the system observables at a set of time points $\{t_l\}$ in condition $\mu$, and (b) the number of non-zero interactions in $W$. Lower cost models tend to have better predictive power than higher cost models.

*Equation 2: Model configuration cost function*

$$C(W) = \beta \sum_{l}^{L} \sum_{i}^{N} \sum_{\mu}^{M} \left( x_i^\mu(t_l) - x_i^{\mu*}(t_l) \right)^2 + \lambda \sum_{i}^{N} \sum_{j \neq i}^{N} \delta(w_{ij}) \qquad 2.a$$

$$\delta(w_{ij}) = 1 \quad \text{if } w_{ij} \neq 0 \qquad 2.b$$
$$\delta(w_{ij}) = 0 \quad \text{if } w_{ij} = 0$$

$C(W)$ is thus the error-plus-complexity cost of a parameter configuration $W$. The cost components are weighted by $\beta$ and $\lambda$, respectively. The complexity cost term is an L0 penalty [45,46,47,48,49] that penalizes non-zero entries in $W$ and is included to both avoid over-fitting and reflect the empirical observation that realistic biological network models such as gene regulatory networks or protein-protein interaction networks are sparse [50], and although L1 penalties are convex and therefore amenable to efficient convex optimization methods, they are often a weaker constraint on the complexity of the interaction matrix $W$ [25,51,52,53]. We do not wish to include direct self-interaction in the present version thus the sum in the complexity term does not include the diagonal elements $w_{ii}$. The computational challenge of network inference is to translate the information contained in a set of experimental observations into an optimal set of models, as represented by a set of low cost interaction matrices $W$. In this report, as we work in the steady state approximation, we ignore the time variable in the cost function in equation 2, and only consider steady state solutions of equation 1 based on two data points at $t=t_0$ and $t=t_s$.

*De novo network inference is a hard problem*

In principle, to infer optimal network configurations one has to compute the cost of all possible network configurations. However, explicit enumeration and cost calculation of all possible parameter configurations $W$ is a prohibitively complicated task for even moderate size systems. To estimate the complexity of this task, assume that any $w_{ij}$ can take on $K$ discrete values out of a value range $\Omega$, for example, $\Omega=\{-1,0,+1\}$ for $K=3$, representing inhibition, no interaction and activation, respectively. As the number of model nodes $N$ increases, the number of possible parameter assignments increases as $K^{N^2}$. Even for moderate $N$, e.g., 20 nodes, the number of distinct solutions is of order $10^{190}$, obviously a very large number, making explicit enumeration prohibitive.

A reasonably clever strategy to traverse this enormous solution space is guided random exploration, e.g., by a traditional Monte Carlo search, in which random moves in multi-dimensional parameter space are kept or rejected based on the cost of the resulting configuration, with a non-zero but small probability of accepting higher cost solutions in order to facilitate the escape from local minima. In an earlier implementation of the network model of equation 1, we successfully used a Monte Carlo search followed by a modified gradient descent method to generate a set of low cost solutions for a relatively small system. This earlier algorithm achieved a reasonable exploration of solution space for a system of 14 nodes, as assessed by the recurrence of dominant interactions across a set of a few hundred low-cost solutions and the agreement of those interactions with well-established knowledge of signaling pathways in cell biology. However, the $K^{N^2}$ argument above and explicit computational benchmarking indicate that such Monte Carlo searches become prohibitively expensive for larger systems.

*Fast inference via a probability model of network configurations*
In search of a more efficient algorithm, we adopted an idea originally developed in statistical physics, and widely used in solving complicated optimization problems in computer science and other areas. Instead of blindly (or cleverly, using a Monte Carlo procedure) sampling an exponentially large solution space by traversing a set of individual solutions (parameter configurations), the idea is to infer system parameters from data in a probabilistic framework. In particular, one first calculates a probability distribution over value assignments for each parameter in the model, and then generates distinct model configurations by sampling from the calculated probability distributions, thus greatly reducing the effective size of the search space.

The basic idea is that models with a large error (or cost) have low probability, while those with a low error (or cost) have high probability. More precisely, the probability of any particular model can be computed from its cost, which depends on its interaction matrix *W* and the experimental data (Equation 2). In statistical physics, there in an analogous relationship between the Hamiltonian for the states of a system and its Boltzmann-Gibbs probability distribution over all states. In Bayesian inference, the below equation relates the posterior distribution of the model on the left to the likelihood function and prior distributions on the right.

*Equation 3: The probability model of network configurations*

$$P(W) = \frac{1}{Z} e^{-C(W)} = \frac{1}{Z} e^{-\beta \sum_{i}^{N} \sum_{\mu}^{M} (x_i^\mu - x_i^{\mu*})^2} e^{-\lambda \sum_{j \neq i}^{N} \delta(w_{ij})}$$

Here, Z is the partition function, which ensures that the sum of the probabilities over all models (all states of the system) is equal to one. In the statistical physics analogy, the exponents contain interaction energies and the parameter β is an inverse temperature (1/T), such that higher values of *β* assign higher probability to lower cost configurations. The parameter λ is the weight of the complexity penalty. The choice of *β* and *λ* is non-trivial and is an open area of research: see methods for more information on how these parameters are set for this study. Given the probabilistic model in equation 3, the practical challenge is to identify parameter sets *W* that represent maximally probable ('good') models, given the data. The explicit computation of probabilities for all possible sets of parameters is not feasible even for moderately sized (N>15) systems. One therefore has to invent practical algorithms for effectively exploring the total solution space and approximately determining sets of good models.

*Iterative optimization of the probability model*

An effective solution is to use an iterative algorithm to approximate the probability distributions of the individual parameters by themselves, often called marginal distributions, from which we can describe high probability, full model configurations. This iterative algorithm begins with a set of random probability distributions. In each iteration step, one assumes approximate knowledge of a set of probability distributions for all parameters ('global information') and then performs optimization updates on individual model parameters ('local updates'). The local updates take effect and become part of the 'global information' for successive iterations, migrating over different individual parameters. The iteration terminates when it reaches convergence to a stable set of probability distributions. At each iteration step, the probability distribution for a local parameter (e.g., a node-node interaction parameter) is calculated based on fitness to experimental data and consistency with the global information about the other parameters. The iterative application of this 'global to local and back' optimization strategy results in probability distributions for all system parameters given a probabilistic model. Such optimal probability distributions are informative in themselves, but also can be used productively to construct a population of explicit individual low-cost parameter configurations, e.g., for explicit simulation of system behavior.

This type of probabilistic method originates in statistical physics and was generalized to a number of hard optimization problems in statistical physics and computer science. An early application of such probabilistic inference was inverse parameter inference for disordered diluted spin systems [54,55,56]. A well known formulation in terms of Bayesian statistics led to the term 'belief propagation' [57].

The belief propagation approach, also known as the Bethe-Peierls approximation or cavity method in statistical physics, provides an approximate method for computing marginal probability distributions on a class of probabilistic graphical models called factor graphs. A joint probability distribution over many variables may factorize into a product of factors. A factor in a factor graph represents an independent contribution to the joint probability distribution, and is connected to the variables that depend on that factor. The BP method is proven to be exact on tree-shaped factor graphs. It has found many useful applications to approximate distributions on sparse factor-graphs [58,59] where the influence of loops in the factor graph is expected to be weak. More recently, several applications to dense, loopy factor-graphs have been proposed [40,60,61,62,63]. The problem we address here is a dense factor graph, where each factor is connected to N model parameters. (Figure 2)

A major advantage of belief propagation algorithms is the reduction of computational complexity. This not only leads to a substantial reduction in computational effort for smaller systems but also opens the door to solving inference problems for larger systems, which would otherwise be prohibitive.

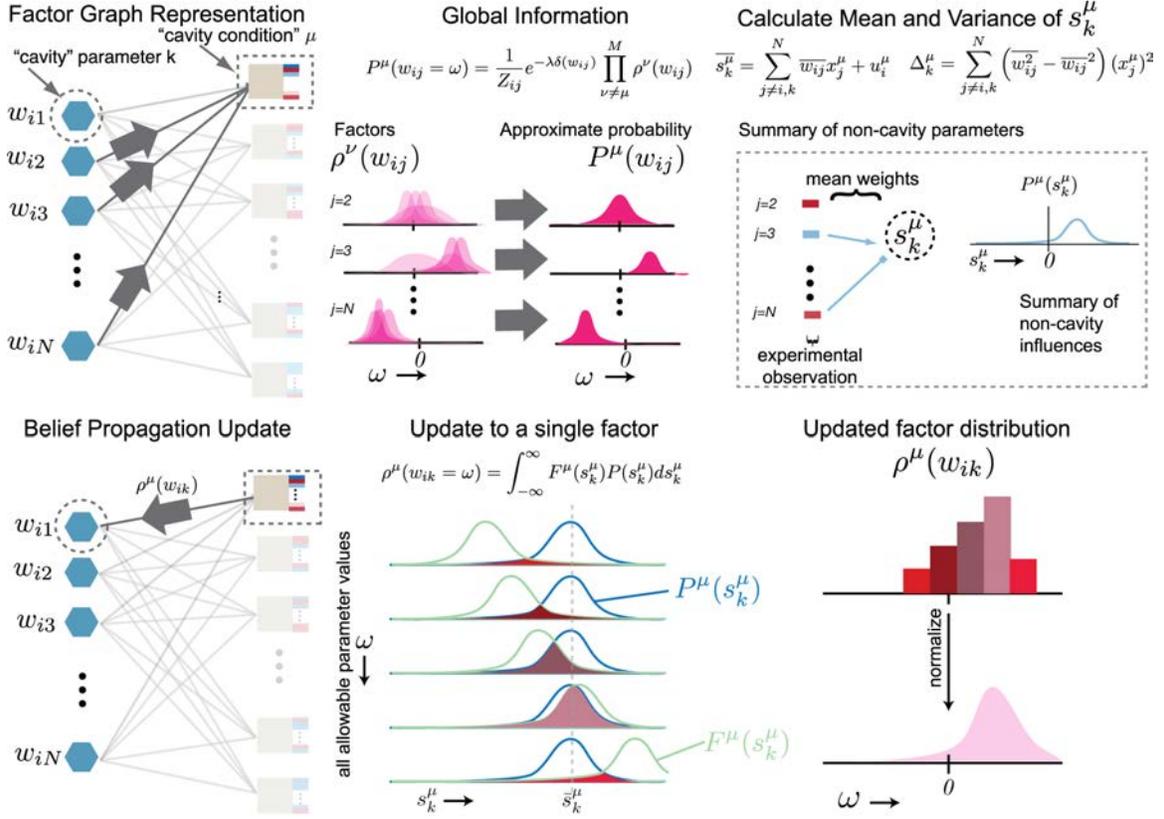

**Figure 2: Iteration process for Belief Propagation.** Top panel: the global information consists of collecting the probability distributions of the non-cavity parameters without the contribution from the cavity condition. This is a simple product over all $\rho^\nu(w_{ij})$ factors except that from the cavity constraint ($\mu$). Distributions centered on zero denote unlikely interactions (see $j=2$), centered on the right of zero denote likely positive interactions (see $j=3$), and centered on the left denote likely negative interactions (see $j=N$). These distributions inform the parameters of the Gaussian distribution for the mean-field, aggregate sum variable $s_k^\mu$. The distribution $P^\mu(s_k^\mu)$ summarizes the state of the non-cavity parameters. Bottom panel: we calculate the probability of each possible parameter assignment ($\omega \in \Omega$) to the cavity parameter ($w_{ik}$) constrained to the data in the cavity condition. This calculation boils down to a simple convolution of the fitness function with a fixed parameter assignment ($F^\mu(s_k^\mu)$) with the probability of the aggregate sum variable $P^\mu(s_k^\mu)$, obtained by integrating over all values of $s_k^\mu$. Each assignment $\omega \in \Omega$ contributes proportional to the area under the curve. The resulting update is the contribution of condition $\mu$ on the distribution of $w_{ik}$, denoted $\rho^\mu(w_{ik})$. This recently updated distribution becomes part of the global information for successive updates to other parameters.

*Simplified probability model of network configurations*

We use a series of assumptions, described below, to simplify equation 3 into a form that can be efficiently calculated without sacrificing the quantitative and predictive nature of the solutions. The assumptions introduced below reduce the problem from a probabilistic description over whole configurations into a collection of probabilistic descriptions for each individual parameter, or 'marginals.' Subsequent sampling from these individual marginal probabilities will result in efficient exploration of high probability model configurations.

**Assumption 1: Discrete set of real valued parameter assignments**. To simplify the probability model, we compute the probability distributions for model parameters over discrete values, from a set $\Omega$, rather than for continuous values. The choice of discretization is important and affects the convergence properties of the BP algorithm and the quality of the resulting probability distributions. Empirically, with the data set at hand we find that 11 discrete values, centered at zero, rarely fails to converge to a stable set of marginal probabilities. Conversely, searching over only 3 weight values results in a high rate of non-convergence. As for the quality of the resulting distributions, the entropies of the marginal distributions are close to zero as we limit the search to only 3 discrete values (see Supplemental Text S4). The entropy is a statistical measure of the uncertainty in the distributions, such that zero entropy distributions imply absolute certainty, where each parameter is predicted to be 100% zero or 100% non-zero. This is an undesirable quality since we want the BP marginal distributions to constrain a set of high probability solutions, not return one parameter set. In practice, discretizing with 11 weights gives intuitively reasonable distributions and a reasonable balance between exploring and constraining solution space.

In the final stage of network inference strategy, we refine the set of discrete valued model parameters to solutions with continuous model parameters using a local gradient descent optimization algorithm [33].

**Assumption 2: Decoupling of nodes at steady state**. In the dynamic model, the system variables $\{x(t)_i\}$ are coupled such that a change to any variable propagates to all others via the time derivatives as in Equation 1. A rigorous way to compute the fitness of a configuration is to simulate a configuration and then compare simulation output to the training data. Such a computation, while feasible in principle, is very costly. An alternative solution is to take advantage of the relationship at the steady state (equation 4) where the time-derivative is equal to zero.

*Equation 4: Model equation at steady state*

$$x_i^\mu = -\phi\left(\sum_{j \neq i}^{N} w_{ij} x_j^\mu + u_i^\mu\right) \quad \text{for all variables } i \text{ and experiments } \mu$$

Equation 4 is a system of self-consistent equations for all nodes $\{x_i\}$. To avoid having to do numerical simulation, we replace $\{x_j^\mu\}$ on the right hand side of Equation 4 with experimentally observed $\{x_j^{\mu*}\}$ at the expense of self-consistency (Equation 5).

*Equation 5: Approximate model equation at steady-state*

$$x_i^\mu = \phi\left(\sum_{j \neq i}^N w_{ij} x_j^{\mu*} + u_i^\mu\right)$$

This approximation has the feature that the model variables become uncoupled from each other: the model predicted value of $x_i^\mu$ depends only on the parameters in the $i^{th}$ row of the interaction weight matrix $W$ and on the set $\{x_j^{\mu*}\}$ of experimentally measured values in condition $\mu$. With this, the posterior probability $P(W)$ can be factorized as a product of independent posterior probability distributions over value assignments to individual rows of the weight matrix. Consequently, we have an independent probability distribution for each row of the interaction weight matrix ($W_i$). These rows describe the effect of the variables $j \in \{1 \ldots N\}$ on a single variable $x_i$, according to Equation 5. The resulting factorized expressions are:

*Equation 6a-d: Probability model of configurations with decoupled nodes*

$$P(W) = \prod_i^N P(W_i)$$

$$P(W_i) = \frac{1}{Z_i}(e^{\sum_\mu^M -\beta(x_i^\mu - x_i^{\mu*})^2} e^{-\lambda \sum_{j \neq i}^N \delta(w_{ij})})$$

$$F^\mu(W_i) = e^{-\beta(x_i^\mu - x_i^{\mu*})^2}$$

$$P(W_i) = \frac{e^{-\lambda \sum_{j \neq i}^N \delta(w_{ij})}}{Z_i}\left(\prod_\mu^M F^\mu(W_i)\right)$$

We introduce the notation $F^\mu(W_i)$ to denote the fitness of the model configuration $W_i$ to the data from experimental μ. It is important to note that the posterior probability distribution in Equation 6d factorizes over the fitness functions, such that $F^\mu(W_i)$ contributes independently to the full probability distribution of that configuration. The probability distribution in 6d, however, does not factorize any further, since each parameter $w_{ik}$ in $W_i$ depends on the other parameters in $W_i$. In order to reach good solutions for 6d without enumerating all configurations $W_i$, we apply an iterative method to infer probabilities for the constituent parameters $\{w_{ij} \; \forall \; j\}$, from which we can sample and enumerate high probability configurations $W_i$.

*Belief propagation algorithm: iterative updates of probability estimates*

As already mentioned above, the belief propagation method consists of randomly ordered updates to the probability distributions for individual parameters, one at time. Updates continue until convergence, when the probability distributions do not change between consecutive updates. We describe the method in detail below for single $W_i$ since the procedure is independent and identical for all rows $i \in \{1...N\}$ due to Assumption 2. The update procedure is schematically diagramed in Figure 2.

**Local updates with global information.** A local update takes place inside an abstract 'cavity', which isolates a single parameter distribution to be updated ($w_{ik}$) and data from a single experimental condition ($\mu$). Each update maximizes the fitness to the experimental data in the cavity condition and compatibility with global information (i.e., probability distributions of other parameters), which evolves as the algorithm iterates. A local update is a change to the cavity parameter's distribution constrained to fit the experimental data and global information. A single distribution is locally updated in one step and becomes part of the global information for updating other distributions in successive steps. This local update is repeated in all possible cavities (i.e. all combinations of parameters and conditions) until global convergence of the distributions is reached.

Just as Equation 6d factorizes over different experimental conditions, so too do the probability distributions for any single parameter. Recall that we are optimizing parameters in a single row of $W$ so for the following equations the index $i$ is fixed. We define $\rho^\mu(w_{ij})$ to be an independent factor that represents the probability distribution of a single parameter given data from experiment $\mu$ and compatibility with other parameters. The BP algorithm begins with randomly initialized distributions for all $\rho^\mu(w_{ij})$. Therefore, the algorithm starts with random probability distributions for all parameters. The initial choice of cavity parameter ($w_{ik}$) and cavity condition ($\mu$) is random. The update of the local $\rho^\mu(w_{ik})$ in a single cavity depends on global information, which is simply a set of approximate probability distributions for the non-cavity model parameters ($P^\mu(w_{ij}), j \neq k$). The approximate probability distribution for each model parameter is calculated at each update as a product of factors ($\rho^\upsilon(w_{ij})$) from all experiments except the experimental condition ($\mu$) in the cavity.

*Equation 7: Global information of non-cavity parameters*

$$P^\mu(w_{ij}) = \frac{1}{Z_{ij}} e^{-\lambda \delta(w_{ij})} \prod_{\upsilon \neq \mu}^{M} \rho^\upsilon(w_{ij}) \quad \forall j \neq k$$

$$\delta(w_{ij}) = 0 \text{ if } w_{ij} = 0$$

$$\delta(w_{ij}) = 1 \text{ if } w_{ij} \neq 0$$

The exponent $\lambda\delta(w_{ij})$ is an independent penalty for non-zero parameter assignments, which encodes prior knowledge that most parameters are zero. This helps prevent overly dense network configurations. The superscript $\mu$ in $P^\mu(w_{ij})$ denotes the exclusion of the contribution from experimental condition $\mu$. In the following step, the distribution $\rho^\mu(w_{ik})$ is updated to fit the data in a single experiment $\mu$ in the context of global information.

**Calculating local update** $\rho^\mu(w_{ik})$. We define $W_{i\backslash k}$ to be a configuration of the $i^{th}$ row of $W$ where the parameter $w_{ik}$ is fixed. The distribution $\rho^\mu(w_{ik})$ reflects the fitness of the parameter with $W_{i\backslash k}$ weighted by the probability of observing the particular configuration ($W_{i\backslash k}$) as in Equation 8.

*Equation 8: Local update to probability distribution of cavity parameter*

$$\rho^\mu(w_{ik} = \omega) = \sum_{W_{i\backslash k}} \left( F^\mu(W_{i\backslash k}) P^\mu(W_{i\backslash k}) \right) \quad \forall \omega \in \Omega \quad \text{8.a}$$

$$F^\mu(W_{i\backslash k}) = e^{-\beta\left(x_i^{\mu*} - x_i^\mu\right)^2} \quad \text{8.b}$$

In the field of optimization algorithms, these equations are sometimes referred to as messages as they communicate information between parameter nodes and factor nodes on the factor graph. Thus, Belief Propagation belongs to a class of 'message-passing' algorithms. It is common to see Equation 7 referred to as messages from the parameter nodes to the factor nodes and denoted $P_{j \to \mu}(w_{ij})$. Similarly, Equation 8 can be thought of as a message update from the cavity factor node to the cavity parameter node, denoted $\rho_{\mu \to k}(w_{ik})$.

**Assumption 3: Independent model parameter distributions.** Equation 8 is the mathematical definition of the factor distribution $\rho^\mu(w_{ik})$. However, a brute force approach for calculating $\rho^\mu(w_{ik})$ as in Equation 8a is computationally prohibitive. The first complication is that computation of the joint-probability distribution $P^\mu(W_{i\backslash k})$ is not possible if we consider the interdependencies of the non-cavity parameters. To circumvent this problem, we assume that in the context of the local update, each parameter probability distribution is independent. Then the joint probability distribution can be calculated as the product of the individual parameter distributions.

*Equation 9: Approximation of the joint distribution for the cavity update*

$$P^\mu(W_{i\backslash k}) = \prod_{j \neq k}^{N} P^\mu(w_{ij})$$

Therefore, Equation 8 becomes Equation 10.

*Equation 10: Local update with factorized probability distributions*

$$\rho^\mu(w_{ik} = \omega) = \sum_{W_{i\backslash k}} \left( F^\mu(W_{i\backslash k}) \prod_{j \neq k}^{N} P^\mu(w_{ij}) \right)$$

This equation is equivalent to the sum-product formulation, which is standard in belief propagation algorithm literature [58]. It is important to note that the assumption that the joint probability distribution is factorized as the product of individual distributions (Equation 9) is exact on tree-shaped factor graphs, but is only an approximation in general. This assumption does not extend beyond the context of the cavity update calculation.

**Assumption 4: Gaussian mean-field approximation.** Another complication in Equation 11 is that a brute force implementation of the sum operation requires enumeration over an exponentially large number of configurations, which in total is $K^{N-1}$. Specifically, we replace the sum over multi-variate configurations ($W_{i\backslash k}$) with an integral over a single scalar variable ($s_k^\mu$). To achieve this, we replace the fitness function's dependence on the multivariate configuration $W_{i\backslash k}$ with a single scalar variable $s_k^\mu$. This transformation is explicitly defined in Equation 11, where the new variable $s_k^\mu$ represents the aggregate contribution of the non-cavity parameters to the fitness.

*Equation 11: Aggregate effect of non-cavity parameters*

$$F^\mu(s_k^\mu) = F^\mu(W_{i\backslash k}) = e^{-\beta\left(x_i^{\mu*} - \phi(s_k^\mu + w_{ik}x_k^{\mu*})\right)^2} \qquad 11.a$$

$$s_k^\mu = \sum_{j \neq k}^{N} w_{ij} x_j^{\mu*} + u_i^\mu \qquad 11.b$$

In addition to this change of variables in the fitness function, we also require a description of the probability of the new variable $P(s_k^\mu)$. Note that the dependence of $s_k^\mu$ on $W_{i\backslash k}$ is through a linear combination of the individual parameters, which by assumption 3 are independently distributed. We invoke the central limit theorem to approximate $P(s_k^\mu)$ as a Gaussian [64]. The mean and variance of this Gaussian are described by the means and variances of the global distributions given by $P^\mu(w_{ij}) \; \forall j \neq k$. Thus, we replace the sum over multivariate configurations with the Gaussian integration of $s_k^\mu$ (Equation 12).

*Equation 12: Gaussian integration of local update to cavity parameter*

$$\rho^\mu(w_{ik} = \omega) \approx \int_{-\infty}^{\infty} F^\mu(s_k^\mu) P^\mu(s_k^\mu) ds_k^\mu$$

The explicit calculation of $P^\mu(s_k^\mu)$ is described in Equations 13a-d.

*Equations 13a-d: Statistical description of mean-field parameters*

$$P^\mu(s_k^\mu) = \frac{1}{\sqrt{2\pi\Delta_k^\mu}} e^{-\frac{(\overline{s_k^\mu} - s_k^\mu)^2}{2\Delta_k^\mu}} \quad\quad 13a$$

$$\overline{s_k^\mu} = \sum_{j \neq i,k}^{N} \overline{w_{ij}} x_j^{\mu*} + u_i^\mu \quad\quad 13b$$

$$\Delta_k^\mu = \sum_{j \neq i,k}^{N} \left(\overline{w_{ij}}^2 - \overline{w_{ij}^2}\right)(x_j^{\mu*})^2 \quad\quad 13c$$

$$\overline{w_{ij}} = \sum_\omega P^\mu(w_{ij} = \omega) \quad\quad 13d$$

**Iteration of update equations.** In summary, the following belief propagation equations are calculated for each cavity update iteratively until convergence.

*Equations 14a-b: Update equations*

$$P^\mu(w_{ij}) = \frac{1}{Z_{ij}} e^{-\lambda\delta(w_{ij})} \prod_{\upsilon \neq \mu}^{M} \rho^\mu(w_{ij}) \quad \forall j \neq k \quad\quad 14.a$$

$$\rho^\mu(w_{ik} = \omega) = \frac{1}{Z_{ik}} \frac{1}{\sqrt{2\pi\Delta_k^\mu}} \int_{-\infty}^{\infty} e^{-\beta\left(x_i^{\mu*} - \phi(s_k^\mu + w_{ik}x_k^{\mu*})\right)^2} e^{-\frac{(\overline{s_k^\mu} - s_k^\mu)^2}{2\Delta_k^\mu}} ds_k^\mu \quad \forall \omega \in \Omega \quad\quad 14.b$$

**Calculation of parameter probability distributions.** When the above iterative equations converge, the final distributions for each of the model parameters are calculated from the set of factors, reflecting the information from experimental constraints.

*Equation 15: Probability distribution of a single parameter.*

$$P(w_{ij}) = e^{-\lambda\delta(w_{ij})} \prod_\mu^{M} \rho^\mu(w_{ij})$$

The belief propagation algorithm provides probability distributions of model parameters characterizing a set of good models. Thus, we reduce the unbounded model search space to a set of tractable probability distributions for model parameters. Next, one must generate high probability models by drawing from the BP calculated probability distributions.

### Network model instantiation by 'decimation' guided by belief propagation

The belief propagation (BP) algorithm generates a set of probability distributions for each model parameter. However, we need distinct model solutions to proceed with predictive and quantitative analysis of signaling pathways. Distinct solutions are generated by the belief propagation guided decimation algorithm [43] from BP generated distributions for each interaction. The decimation algorithm works as follows. (i) An initial BP is run to compute probability distributions $P(W_{ij})$ for all possible interactions. (ii) A possible interaction (suppose an edge that connects node k and node l) and an associated edge value ($w_{kl}=\omega$) is chosen. The edge is chosen based on a probabilistic draw among all possible edge values for all edges. The probability of choosing an edge value is proportional to its BP computed probability. (iii) The probability for the chosen edge value is fixed to $P(W_{kl}=\omega)=1$. (iv) A second round of BP is run with fixed $P(W_{kl}=\omega)=1$. (v) Steps i-iv are repeated until an edge value is fixed for all possible interactions in the system. This procedure generates thousands of network models with varying configurations and error profiles. The non-zero parameters in each model are further optimized using a gradient descent algorithm, which relaxes the discretization of parameter values and further lowers the error by fine-tuning the real number values of the parameters. Moreover, the gradient descent refinement ensures that the network models are mathematically in steady state and the nodes in the network models are fully coupled [33]. Each model is then a set of differential equations describing the behavior of the system in response to perturbations.

**Direct vs. Indirect Interactions**: Due to limitations in experimental measurements, not all proteins and their many phosphorylated sates are directly measureable. The result is that many key intermediate players are excluded from the model. For this reason, direct interactions in our model do not necessarily imply direct biological interactions (Supplemental Figure S15). Increasing the number and quality of protein and phospho-protein measurements increases the chance of modeling direct interactions.

# Technical performance

We developed an experimental-computational technology, in the spirit of systems biology, which integrates perturbation experiments and network inference algorithms to construct quantitative models of signaling pathways and predict responses to drug combinations in a collection of cells. We first tested our technology to infer network models of artificially generated data. With these technical tests, we demonstrated that we can infer network models with good accuracy and achieved major speed performance improvement over conventional Monte Carlo (MC) based methods. Furthermore, the synthetic data is generated without the assumptions utilized in the formulation of BP, and therefore serves as a reasonable test of the sensitivity of the BP method to those assumptions. Next, we applied the technology to perturbation data in a RAF inhibitor resistant melanoma cell line (SKMEL-133). We demonstrated that the technology successfully captured already known interactions and predicted new ones in signaling pathways. We tested the quantitative descriptive power of the method by quantifying the ability to reproduce the response profiles from withheld experiments. We then used *in silico* simulations and to predict the effects of novel drug perturbations, particularly the inhibition of Polo-like Kinase 1 (PLK1).

The success of BP depends on whether or not the simulations of the models taken from BP are quantitatively predictive of cellular response to drug combinations. It is also useful, however, to evaluate the overall performance of the BP algorithm on data sets engineered from completely known networks.  With such toy datasets we (i) demonstrated that BP converges quickly and correctly,  (ii) compared BP network models to the known data-generating network, and (iii) evaluate performance in biologically realistic conditions of noise and sparse perturbations.  See methods for more information on how the toy data is generated.

**Belief Propagation is fast and accurate**.  MC is a strategy for sampling the space of explicit solutions, in which full parameter configurations are searched as a whole.  Short of infinite coverage, a thorough MC search yields reasonably accurate approximations of the true probability distributions; both posterior probability distributions of explicit configurations, and marginal probabilities of individual parameters, calculated by counting the frequency of any parameter assignment across the set of good solutions. MC is a ubiquitous optimization strategy in the statistical physics community, and its ubiquity makes it a valuable candidate for comparison. We examined speed and accuracy performance of MC and BP for increasingly large models.  To do this, one toy data generator was constructed for each of the ten different sizes from $N=10$ to $N=100$.  The number of training patterns equaled the number of nodes for consistency of comparison, i.e. $M=N$.  Both methods searched a very large parameter space of 41 possible parameter assignments with $\Omega=\{-2,-1.9,\ldots,1.9,2.0\}$; thus for this toy dataset the search space of all configurations was of order $41^N$.

The first criterion of interest was time of convergence (Figure 3A). For both methods, the time required for convergence increases as the size of the system increased, but MC was consistently about three orders of magnitude slower than BP. The speed advantage of BP is vital for our ability to scale up the size of *de novo* model construction to hundreds of proteins, phospho-proteins and cellular phenotypes.

The second criterion was accuracy. While we were interested primarily in the accuracy of predicting responses to new perturbations, we were also interested in the accuracy of the inferred interactions as an indicator of the models' explanatory and predictive power. In practice, we found that BP did no worse than MC on these datasets given reasonable termination conditions for MC (Supplemental Text-S5).

A Pearson correlation coefficient between the set of non-zero parameters in the data generators and the average values inferred by BP (blue) was a reasonable measure of agreement between true and inferred parameters.  BP resulted in a correlation of $R=0.7$, which was quite high considering the relatively small number of training patterns used for each inference ($M=N$). As the number and quality of training pattern increases, the correlation approached 1 (Supplemental Text S1, Supplemental Figures S2 and S3).  A more critical metric of accuracy was the agreement with the training data.  The average parameter strengths inferred from BP and MC were used to calculate the expected value of the data points (as in Equation 6D).  Both MC and BP reproduced the training data very well as quantified by mean squared error per data point (Figure 3C). For reference, both BP and MC errors were at least two orders of magnitude better than what was expected at random.

While other parameter search methods such as genetic algorithms [65], simulated annealing [66], regression [52,67], lasso-regression [52], Hybrid Monte Carlo [68,69,70] and Kalman filtering [71] may offer improvements in speed over the standard MC comparison used here, they also suffer from poor scaling properties in the absence of prior knowledge. A statistical-mechanics analysis demonstrated that BP outperforms mutual information based inference algorithms since BP takes into account the collective influence from multiple inputs [72]. We conclude that BP offered a tremendous speed advantage over MC at no observable loss of accuracy.

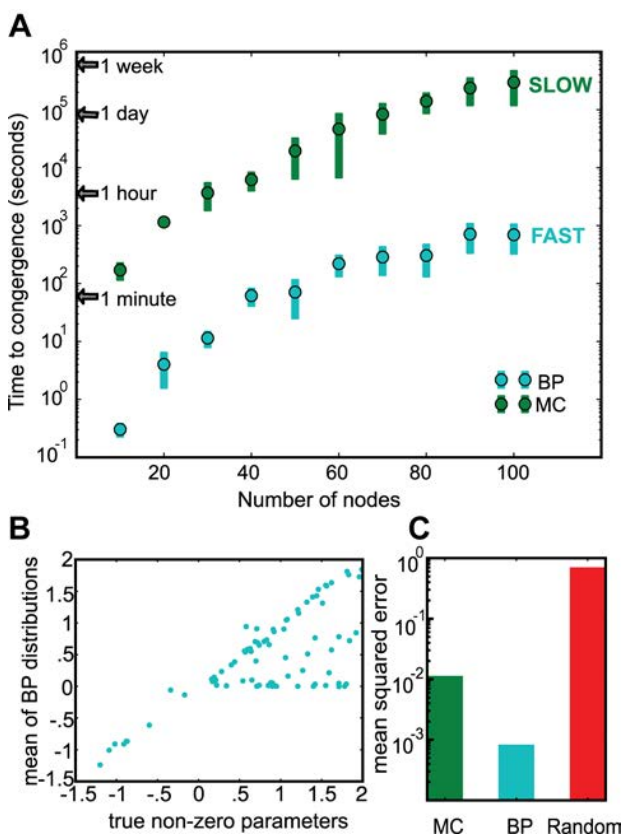

**Figure 3: BP is significantly faster than MC with comparable accuracy.** (A) BP converges three orders of magnitude faster than MC, even as the size of the system increases to 100 nodes. In this test, the number of training patterns equals the number of nodes in both BP and MC. (B) The means of the distributions from BP are plotted against the true non-zero parameters from the set of the data generators. BP has a high correlation (R=0.7) with the true parameter values, with many points exactly on the diagonal. (C) MC and BP produce low errors per data point compared to random interaction assignments (Red bar).

**BP reproduces true interactions.** BP inference was fast and almost perfect when the system had been sufficiently explored by perturbations. In the case of toy data, one can perturb any set of nodes simultaneously with complete control, and generate information-rich data sets. Use of rich data sets provided a sufficient training set for BP to produce near perfect inference of the underlying system (Supplemental Figure S2 and S3). In biological experimental conditions, however, we are limited by the availability, strength and specificity of the drugs, by the availability of reporters (such as antibodies) as well as the technical accuracy of the measurements. We are further limited by the financial and temporal cost of testing all combinations, even for the drugs that are available.

Here, we evaluated BP performance in biologically realistic conditions; small number of sparse perturbations applied individually and in pairs. The inference is repeated with added noise to evaluate sensitivity to noise. The Gene Network Generator GeNGe [73] constructed the structure of positive and negative interactions for the data generator. The data generator network contained several common regulatory motifs, including feedback loops, single/multiple input motifs, multi-component loops and regular chains. For this study, a drug was represented as a strong inhibition of a main target and smaller positive or negative effects on four or fewer other nodes, which were meant to simulate off-target effects. Complete knowledge of the perturbations and off-target effects was used for the noise-free results, while only knowledge of the main targets was used for the noisy data results thus simulating inference on drugs with unknown off-target effects. We simulated the system to steady state in response to each condition of 14 *in silico* drugs applied individually and in pairs. The steady state profiles were recorded and used as the training data, while those conditions that oscillated were excluded.

Ultimately, the predictive power of the inferred models can only be assessed by explicit simulation of individual models and comparison with experiment. However, the average value of the BP inferred probability distributions can be used in a descriptive sense and either guide human intuitive understanding of biological pathways or be compared to prior knowledge.

The performance features for evaluating the inferred interactions were recall and precision. Recall is the fraction of interactions from the data-generating network that were correctly inferred in BP generated probability distributions. False negatives decrease the recall fraction. Precision is the fraction of interactions inferred from probability distributions that were also in the data generator. False positives lower the precision fraction.

The average interactions as derived from belief propagation yielded a sparse network with a significant number of true interactions (Figure 4A). Importantly, some incorrectly inferred interactions were not mutually exclusive of correctly inferred interactions. That is, since each row of $W$ was inferred independently (Assumption 2), one might expect a row to be either all correct or all incorrect, yet many rows in Figure 3B had both true and false inferences.

Interestingly, false positives and false negatives are somehow structurally related. In other words, BP tended to miss one or more true interactions (false negatives) and replace them with one or more compensatory interactions (false positives) that are structurally adjacent to the missed interactions. We observed three common structural correlation motifs, which we refer to as (a) upstream, (b) symmetric, and (c) co-regulation motifs (Figure 4B).

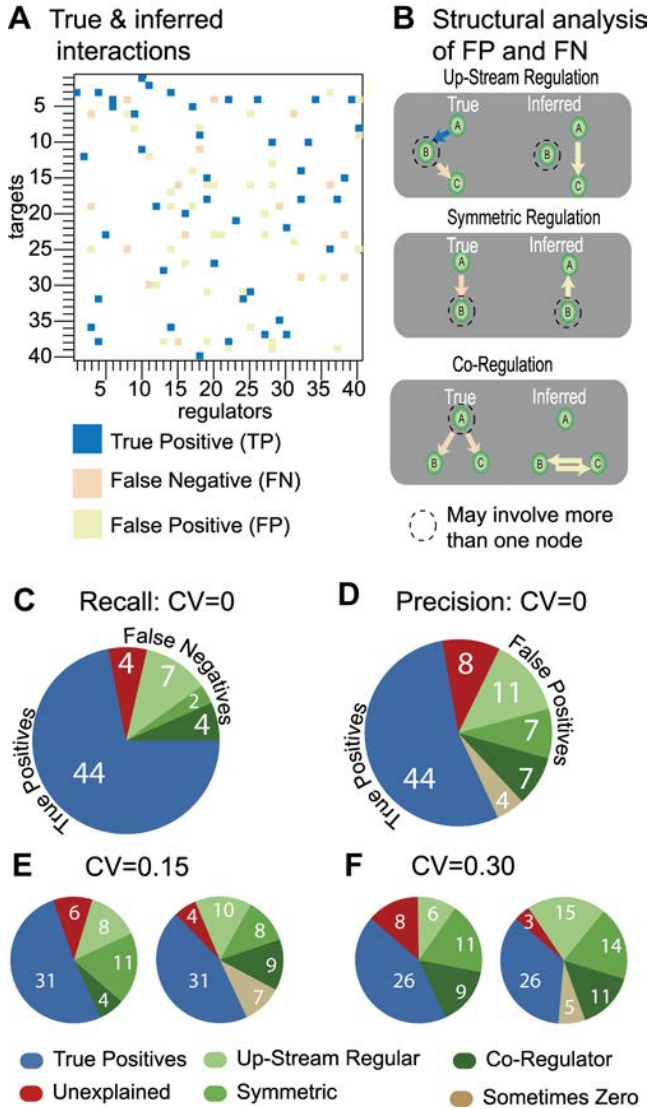

In the upstream motif, nodes A and C are connected through an intermediate node B, but an edge is inferred from A to C directly. In the symmetric motif, a false positive connects two directly connected nodes but in the wrong direction. In the co-regulation motif, node A directly regulates B and C separately, but a false positive exists between B and C directly. In addition to being structurally correlated, the numerical correlation between nodes involved in false positives were observably high in the training data (Supplemental Figure S1).

BP missed 17 of the 60 true interactions, giving a recall of 74%. Only 4 of these 17 false-negatives were not involved in one of the three structural correlation motifs. Meanwhile, BP predicted 37 false positive interactions for a precision of 55%. However, 29 of these were either involved in one of the three motifs or had a significant probability of being zero (and therefore ceasing to be false-positives). Consequently, we concluded that while many of the false positives may seem worrisome, they were supported in the data and in the underlying data-generating network.

The results on this toy data confirmed that this implementation of BP had trouble disambiguating correlation and causation from steady-state data, which is a difficulty common in analysis of steady-state data [74]. BP inferred interactions between highly correlated nodes even if they were not causally connected. BP was better able to infer causality when there was sufficient perturbation of the nodes involved in a potential interaction.

**Figure 4: Detailed performance on a single synthetic data-generating network.** The average parameters from the BP distributions are compared with the true interactions in the synthetic data generator. The color-coded matrix (A) summarizes all inferred and true interactions. While BP recovers many of the true interactions, some of the interactions are missing (orange; false negatives) while others are incorrect (yellow; false positives). We identified three compensatory motifs (B), which relate false positives to false negatives. Collectively, these classes of compensatory motifs contribute to most of the false negatives (C) and false positives (D). In D, we've also included a category for interactions that have a significant probability of being zero (a non interaction). Even in the presence of considerable noise, (E, F) a significant number of interactions are correctly captured and most of the falsely inferred edges participate in compensatory motifs.

It is likely that the assumptions inherent in this BP algorithm may be the root cause of the incorrect edge predictions, in particular, assumption 2, which decouples the likelihood function from the dynamics of the system. We expect that combining a tailored likelihood function to incorporate time-series data may dramatically improve the ability of a similar BP method to infer causality more efficiently.

**The inference of network parameters was moderately sensitive to noise.** With toy models, we could accurately analyze the effect that noisy data had on the accuracy of inferring network interactions. Noise from measurement technology can have deleterious effects on network inference and introduce sensitivity to data outliers. For example, RPPA produces Gaussian distributed data in the absence of substantial biological variability [75]. We estimated a coefficient of variance (CV) of 15% on the measurements from RPPA. To examine the effects of Gaussian noise on BP inference, we applied Gaussian distributed noise ($G_{0,\gamma}$) with a mean of zero and standard deviation of $\gamma$ representing the CV as in Equation 16.

*Equation 16*

$$x_i^\mu = x_i^\mu \left(1 + G_{0,\gamma}\right)$$

We constructed two data sets with Gaussian noise; one with a realistic CV of 15% ($\gamma = .15$) and one with high CV of 30% ($\gamma = .3$) as a worst-case scenario. Though both recall and precision decreased with added noise (Figure 4E, 4F), the number of unexplained interactions stayed roughly constant. Importantly, BP was still able to identify key regulatory influences from the noisy steady-state data from sparse perturbations without any dependence on prior knowledge. The performance on noisy data approached the performance of noise-free data when sufficient numbers of replicates were used in the training.

This analysis of BP inferred interactions was limited to a thorough examination of a single set of interactions, taken as the average of each parameter from the BP generated distributions. We have demonstrated that BP is fast, accurate and minimally sensitive to realistic amounts of noise. Moreover, BP was sufficiently strong in distinguishing causal from correlated relationships even though we are currently limited to steady-state data from a small set of perturbation conditions. We know how good the inference of interactions is in a scenario where a perfect model of the data generator exists. The comparison gives us an idea of the structural plausibility of the interactions inferred in real biological contexts.

# Network models of signaling pathways in melanoma cells

The performance and accuracy of the BP algorithm in reconstructing artificial, biologically inspired models from toy data sets suggest that the BP algorithm may work for inferring useful predictive signaling network models from systematic perturbation experiments. The probabilistic nature of the BP algorithm is the key feature that enables de novo (without prior knowledge) inference on large and complex problems of cell biology, e.g., signaling processes involving more than a hundred molecular and phenotypic variables, which previous methods could not reach. Given this opportunity, we have applied this network pharmacology approach to SKMEL-133, a melanoma cell line resistant to a RAF inhibitor (Vemurafenib, PLX4032).

*Experiments: systematic perturbation of SKMEL-133 cells using drug pairs*
We systematically perturbed SKMEL-133 cells with a panel of targeted drugs (Figure 5) used singly and in paired combinations. The cell lines were systematically perturbed with 44 dual or single perturbations using combinations of 8 different drugs (Figure 5). The drugs were selected on the basis of their target specificity, availability and usability in clinical settings. The selected drugs predominantly target the PI3K/AKT and MAPK pathways, which are known to affect the response to RAF and MEK inhibitors in some melanoma cell lines and clinical samples [76,77,78,79,80]. Drug concentrations were chosen based on the effect of each single drug on a likely downstream effector of its target, e.g, as assayed using Western blot (Suppl. Figure 4). The changes to potential targets were measured with Western blot experiments in different drug concentrations and standard response parameters, such as IC50 values were determined from the dose-response curve. (Supplemental Figure S4). For example, AKT inhibitor (AKTi) concentration was chosen based on reduction in AKT phosphorylation at S473 as measured by using Western Blot. The drug dose response curve indicated that ~5000nM of AKTi was required to reduce phosphorylation by 40%. As a compromise between the competing requirements of gentle perturbations and observable effects, drug concentrations, at which the downstream effect of a single perturbation was reduced by 40% relative to the unperturbed control were used in the paired perturbation experiments. The SKMEL-133 cell cultures were incubated for 24 hours upon drug perturbation to ensure a measurable physiological response interpretable as a steady state (details in Methods). The main intent of the systematic perturbations was to explore diverse aspects of the signaling response and for the experimental readout, such as protein concentrations, to be maximally informative about the couplings in a network model.

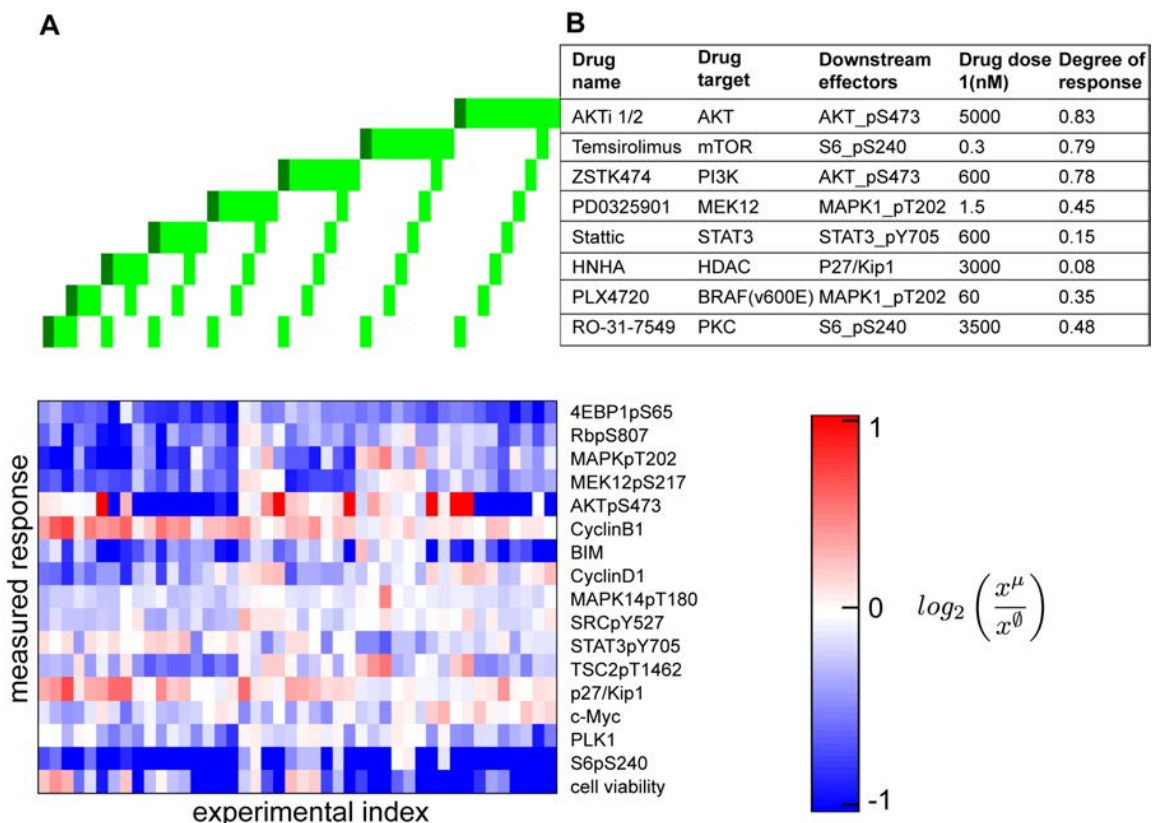

**Figure 5: Systematic perturbation experiments**. (A) Perturbation experiments with systematic combinations of eight small molecule inhibitors, applied in pairs and as single agents in low (light green) and high (dark green) doses. The perturbation agents target specific signaling molecules. (B) The response profile of melanoma cells to perturbations. The response profile includes changes in 16 protein levels (total and phosho-levels, measured with RPPA technology) and cell viability phenotype relative to those in no-drug applied condition.

### Experiments: observation of response profiles in SKMEL-133 cells

After inhibition of one or two targets, we assessed the response to perturbation for each protein node $x_i$ by comparing the (phospho)-protein level at the presumed steady state to the level measured in identical experiments in untreated cells; and, similarly for cellular outcomes such as viability. We used an array technique (reverse phase protein arrays, RPPA [81]), in which cell lysates were interrogated by antibodies against proteins of interest. Compared to Western blot assays, RPPA had the advantage of higher throughput, higher sensitivity, better dynamic range but the disadvantage of lack of size separation, which was addressed by using antibodies carefully tested for specificity. These were crucial advantages for the quantitative inference of network models. For each experimental condition, i.e., for each perturbation, three biological replicates were measured. The (phospho)-protein levels were quantified from spot intensities on the protein arrays and log normalized with respect to the corresponding protein levels from non-perturbed control experiments (response=log$_2$(perturbed/unperturbed) ). Viability of the cells after drug perturbation was measured using a resazurin assay 72 hours after the cells were perturbed. Overall, log ratios for 16 protein intensity readouts and the cell viability phenotype in 44 perturbation conditions constitute the training data for *de novo* network inference, from a series of experiments using all combinations of 8 drugs (28 experiments) and 2 drug concentrations for single drug perturbations (16 experiments), (Figure 5). Cell viability measurements at 72hrs were used to ensure the phenotypic responses reached to steady state and significant phenotypic response was observed as a consequence of changes in relatively early proteomic responses to drug perturbations. Indeed, analysis of cell viability changes at 0, 24, 72 and 120 hrs after drug perturbations revealed no significant changes in cell viability at 72 and 120 hrs, whereas the cell viability change at 24hrs was not at steady state (data not shown).

### *Concept of network models of signaling in SKMEL-133*

The proteomic and phenotypic (i.e., cell viability) response profiles to drug perturbations serve as experimental constraints for the construction of network models. We build network models of signaling pathways in the melanoma cell line SKMEL-133 to predict the response to inhibition in single and multiple protein targets. In the current implementation of this approach, non-linear differential equations of the system is defined by a set of nodes (scalar variables), such as molecular concentrations of proteins and phospho-proteins or scalar measures of cell behavior (such as viability, cell death or motility). The edges connecting the nodes represent the strength of potential (logical or physical) interactions between two molecular levels and between molecular levels and cellular outcome variables. We mathematically represented the network model by a set of ordinary differential equations, which describe the time behavior of the system following a perturbation from a well defined initial state. The edges are the system parameters to be optimized for maximal predictive power. Each measurement represented a constraint in the optimization algorithm. After derivation of optimal interactions parameters, the resulting network models quantitatively linked proteomic changes to each other and to phenotypic changes. We expected the network models to generate hypotheses about previously unidentified interactions, some of which may be accessible to biochemical experimental tests; and, to provide a quantitative, as well as intuitive, model of the signaling cascades in melanoma cells and to be useful for the prediction of nontrivial drug combinations (or single drugs) that may be effective in reducing cell proliferation and, in future work, in overcoming or preventing the emergence of drug resistance. The details of the mathematical model and the inference strategy are described in the Theory section.

### *Leave-k-out tests of predictive power of network models*

In order to test the accuracy and the predictive power of the models, we used a leave-k-out validation strategy. For each leave-k-out test, we withheld data from k experiments, inferred network models on the remaining experimental data, predicted molecular responses corresponding to those in withheld experiments through numerical simulations and compared the predictions with the experimental observations in the withheld experiments. Specifically, for the SKMEL-133 cell line, network models were inferred using a set of perturbation experiments that includes paired combinations of 7 drugs + 1 drug applied as a single agent in two different doses so that 7 unique combinations of the single agent are withheld. For each partial dataset, we generated 1000 network models with the BP guided decimation algorithm. Overall, eight sets of network models were generated by withholding experimental information for all possible combinations with one drug at a time, but using data from all single drug experiments. The network models were numerically simulated using the model equation 1 and virtual perturbations mimicking the perturbations in withheld experiments to predict the proteomic and phenotypic responses to drug combinations. Response trajectories in time were generated for each read-out until the system reaches steady state and the predicted steady state response was computed by averaging the simulated steady state values over network models, using even weights for all 100 models with lowest error, for simplicity. Comparison of the predicted steady state response profiles for each condition with the values in withheld experiments indicated good agreement between the experimental and predicted response profiles (Figure 6). The correlation between the experimental and modeled response profiles at steady state was chosen as a simple measure of predictive power. The comparison resulted in an overall Pearson correlation coefficient of 0.87 (cross-validation error CV=0.05; see methods for definition of CV) between experimental and predicted proteomic and phenotypic measurements (Figure 6, Supplementary Figures S7 and S8), which we considered good to excellent.

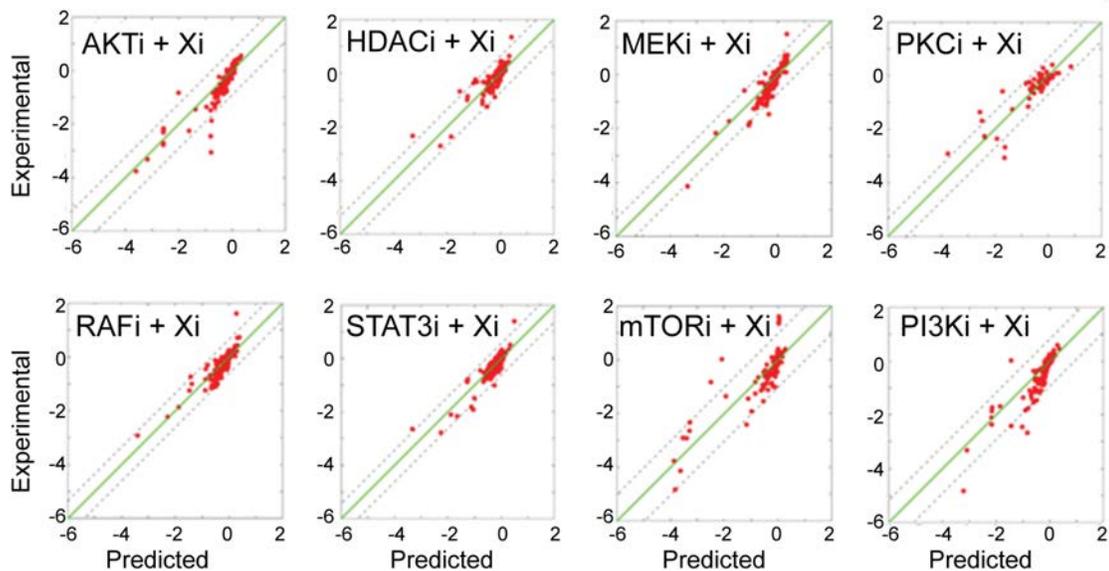

**Figure 6: Predictive power of network models.** Eight distinct leave-7-out cross validation calculations indicate a strong fit between the predicted and experimental response profiles. In each cross-validation experiment, network models are inferred with partial data, which lacks responses to all combinations of a given drug. Next, network models are executed with *in silico* perturbations to predict the withheld conditions. The cumulative correlation coefficient in all conditions between predicted and experimental profiles is 0.87 (CV=0.05). Few prediction outliers deviate from experimental values more than 1 σ (standard deviation of the experimental values, dashed lines).

*Interpretation and predictions from de novo inferred network models in melanoma cell lines*

Using our network inference algorithm, we derived a set of quantitative network models from paired perturbation experiments on SKMEL-133 cells. The network models have 25 nodes, i.e., 16 levels of proteins or phospho-proteins (phosphorylated on specific sites as indicated), one phenotype node corresponding to cell viability measurements and 8 activity nodes, which quantify enzymatic activity.

**Average network model as a guide.** For simplicity of visualization, rather than attempting to visualize many individual models, a model representing averaged interactions, called the 'average network model', was constructed (Figure 7B) by averaging over the interactions in the set of 100 network models with lowest model errors among the 1000 instantiated models. Note that no quantitative prediction can be derived by simulation from the average model (see above for predictive power of the ensemble of solutions). However, the average model depicted the qualitative features of the solution space by reporting, for the most part, interaction values that were present at high frequency and amplitude in the solution space. Remarkably, as summarized in the average model, the network solutions captured the signaling interactions within the major pathways important for melanoma progression, e.g., well-accepted MAPK or PI3K/AKT pathways. In addition, the average model was suggestive of a series of additional interactions (see Supplementary Table S1 for details), some of which may represent novel biochemical interactions and may be good candidates for experimental tests ('validation').

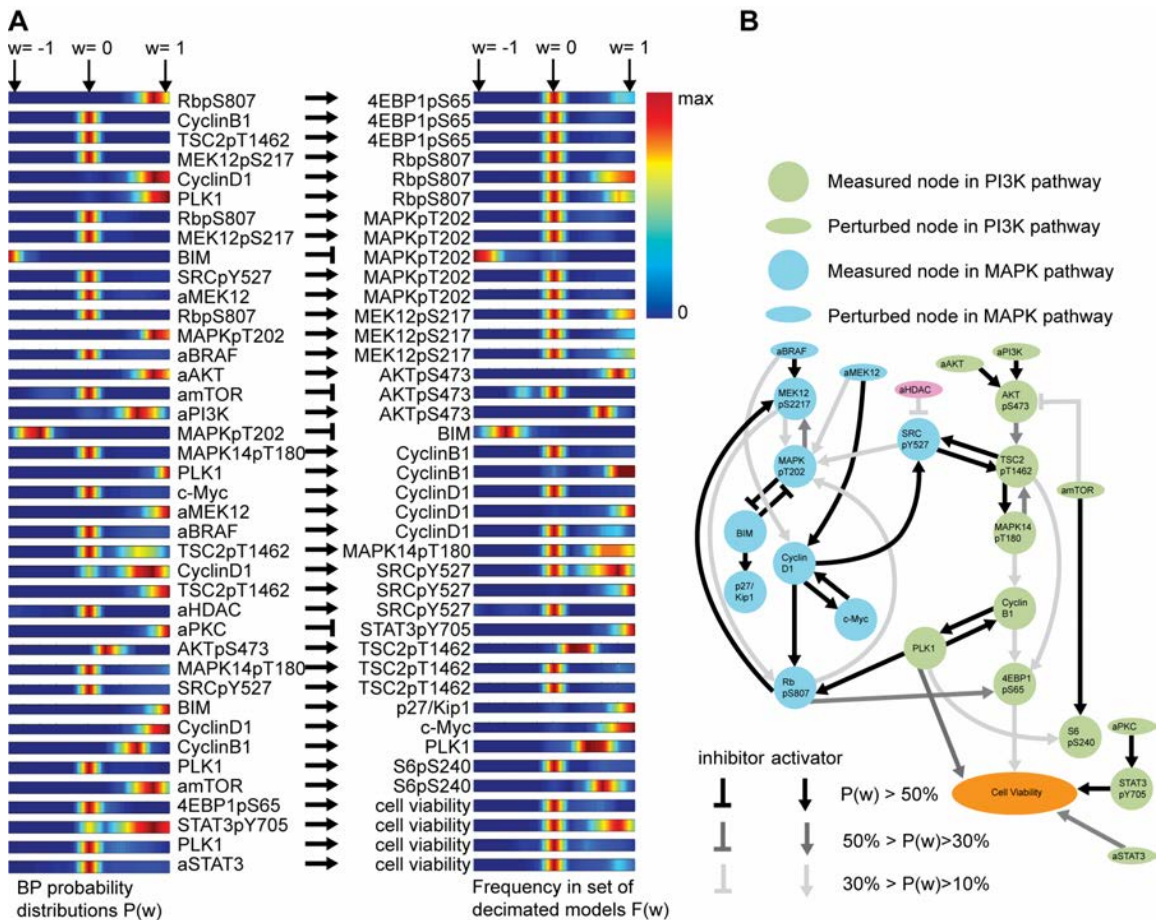

**Figure 7: The distribution of edges in all network models and average network model**. The probability distribution of edge values ($w_{ij}$) after BP (A, left) are similar to the histograms of the corresponding interactions after decimation (A, right). An interaction strength is nonzero when it has high amplitude and frequency in solution space. We generate instantiated models with BP guided decimation algorithm followed by gradient descent optimization. According to the agreement between the distributions in two panels, BP probability distribution and final model histogram are similar to each other with important exceptions. Thus, BP guided decimation algorithm goes beyond simply sampling from the BP models and introduce features such as mutual exclusivity to the final models. (B) The average network model over best solutions (100 models) with lowest errors. The opacity of the edges scales with the absolute strength of the edges. The network diagram is shows the edges that have the highest amplitudes and frequencies in the solution set. The known interactions such as those in RAF/MEK/MAPK and PI3K/AKT pathway are captured by the inference method.

**Interpretation: AKT pathway.** The canonical (known) PI3K/AKT pathway is characterized by a series of complex interactions resulting from the activity of the upstream kinase PI3K and reaching more downstream regulatory proteins [82]. Although the detailed spatiotemporal regulation of the PI3K/AKT pathway and its phenotypic output ranging from proliferation to metabolic changes are highly complex, our network models captured the major known interactions in this pathway. Our models described the pathways such as the AKT pathway only in terms of the observed nodes. The details of the models deviated from the intricate details of the known pathways due to the existence of unobserved nodes. As a natural consequence, the predictions that can be extracted from these models were also limited to the scope of the observed nodes and inferred model variables for those observed nodes. With the expansion of the observed nodes and perturbations, the network models may converge to the detailed pathway descriptions with additional strength of quantitative predictions.

The indirect positive effect of PI3K on AKT phosphorylation, phosphorylation of TSC2 on T1462 by AKT activity, the regulation of S6 and 4E-BP1 by the PI3K/AKT pathway were all represented in the inferred network models. In our model, mTOR activity (node amTOR, where a stands for activity, see Methods) has no allowed upstream node. This is due to our modeling approach, where activity nodes are quantified by the strength of the perturbations acting on them and to our experimental scheme, where no direct proteomic measurement is collected for the activity nodes (See methods for details). However, the edge connecting TSC2pT1462 to 4E-BP1pS65 links the upstream components of the pathway to the downstream components and compensates for the lack of inferred direct interactions upstream (i.e. PI3K/AKT) to downstream components (i,e, mTOR) and leads to changes in downstream elements upon perturbation of the upstream components of the pathway. An inhibitory edge from mTOR to AKTpS473 in the PI3K/AKT pathway is reminiscent of the reported feedback loops in this pathway [83].

**Interpretation: MAPK pathway.** In the RAF/MEK/ERK pathway, the inferred network models captured many of the known interactions that link the MAPK activity to CyclinD1 levels and RB phosphorylation [84]. However, the edges that link this pathway to cell viability were indirect through 4E-BP1 phosphorylation and had a relatively low effect on this phenotype (See below for a quantitative analysis). The inferred networks had a bidirectional interaction between PLK1 and CyclinB1 [85]. A strong direct interaction between PLK1 and cell viability was also in the models, consistent with the fact that PLK1 and CyclinB1 regulate the G2 to M transition in the cell cycle and PLK1 has multiple roles in mitosis such as spindle formation [85].

**Prediction: logical and biochemical interactions in the melanoma cell line.** In addition to confirmation of known biological interactions in the *de novo* derived models, the average model also indicates a series of potentially novel interactions. First, a strong bidirectional edge was observed between the phosphoprotein levels RBpS807 and MEKpS217 ($<w_{RBpS807 \rightarrow MEKpS217}> = 0.59$, $<w_{MEKpS217 \rightarrow RBpS807}> = 0.19$). A direct interaction between RAF and Rb, phosphorylation of Rb by RAF and apoptosis induction upon disruption of this interaction have already been shown in melanoma [86,87]. However, the regulation of MEK phosphorylation by the inhibitory phosphorylation on Rb had not been reported in melanoma cells. Interestingly, analysis of pituitary gland tumor formation in Rb and K-RAS double-knockout mice had suggested a genetic interaction between Rb and RAS, an indirect upstream regulator of MEK [88,89]. Moreover, in cell culture experiments, inactivation of Rb in murine fibroblasts resulted in aberrantly elevated RAS and MAPK activity, which was consistent with the positive edge observed between RbpS807 and MEKp217 [88,89]. A detailed molecular biology and genetic study is needed to determine whether the observed Rb/MEK bidirectional interaction is biologically relevant in melanoma or incorrectly modeled. Second, an inhibitory edge from the HDAC activity node to SrcpY527, a critical phosphorylation site for the auto-inhibition of Src [90] was consistent with the direct and indirect interactions of HDAC isoforms with Src observed in multiple cancer contexts [91,92]. Finally, positive edges connecting CyclinD1 and TSC2pY1462 to the SrcpY527 node were in the average model. No such interaction, nor any interaction that can compensate the observed edges, have been reported, to the best of our knowledge. The observed edges may correspond to logical interactions, i.e., via intermediate physical interactions not explicitly in the model, that coordinate a negative feedback loop acting on Src from the MAPK and PI3K pathways. All of the predicted interactions are suggestive of follow-up genetic and/or biochemical experiments.

**Prediction: target identification in the melanoma cell line via computation of viability changes in response to untested perturbations.** Once we had a set of network models of SKMEL-133 cells, we could predict the effect of arbitrary *in silico* perturbations acting on any protein node (concentration or activity) present in the model. Such *in silico* perturbations can be used to predict the quantitative effects of hypothetical drugs (i.e. virtual drugs that target the nodes in a model) on the cellular phenotypes; and, to quantitatively track signaling processes, beyond the qualitative interpretation of the average network model (Figure 6-Right).

As an example, to nominate novel drug targets, we predicted the phenotypic (i.e. cell viability) response of SKMEL-133 cells to various perturbations by executing the model equation 1 with the parameters *W* determined from the experimental information. The simulations, for each of the 1000 network model solutions, yield the time progression of the concentrations of molecular species. In each simulation, a single perturbation was applied to a single protein node that reduces the activity of that node to 50%. The analysis of the phenotypic responses at steady state revealed that (separate) perturbations of 4 particularly interesting nodes, i.e, potential drug targets, resulted in the most significant alterations in cell viability (Figure 8).

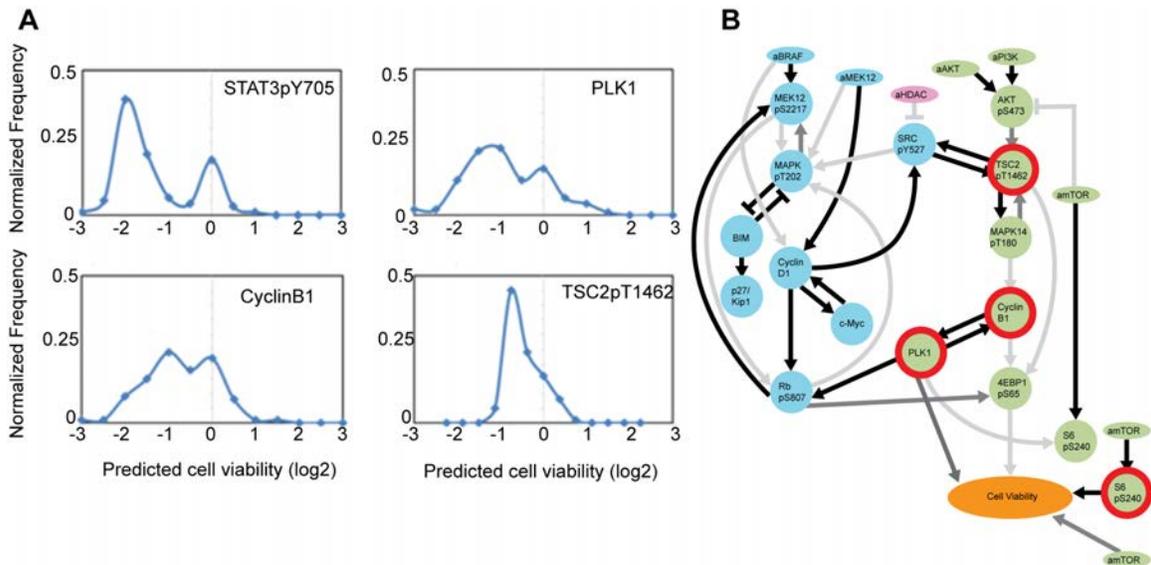

**Figure 8: Novel predictions with in silico perturbations**. (A) The histogram of phenotypic response profiles to the four most effective virtual perturbations from the best 100 network models. The response to STAT3p705 reflects the effect of PKCi on cell viability. Viability changes in response to perturbations on cell cycle proteins PLK1 and CyclinB1 are genuine predictions from the network models. Perturbation of TSC2pT1462 (inhibitory phosphorylation) down regulates the PI3K/AKT pathway and leads to a decrease in cell viability in the PTEN null SKMEL133 cell line. (B) The perturbed nodes that lead to reduction in cell viability in the context of average network model (Circled in red).

First, the most significant reduction in cellular viability was predicted in response to perturbation of STAT3pS705 (average log(relative cellular viability) = -1.16). In the network models, the STAT3 phosphorylation node was downstream of the PKC activity node and our perturbation experiments did show that PKC inhibition lead to low cell counts. Thus, the response of STAT3 inhibition reflected the effect of PKC inhibition, which was experimentally shown and used as an input in the training set. Next, perturbation of both PLK1 and CyclinB1 protein nodes (total protein level) lead to significant reduction in cellular viability (average log(relative cellular viability): -0.85 and -0.69 respectively). PLK1 and cyclinB1 are two important cell cycle proteins regulating G2/M transition and highly specific PLK1 inhibitors are potential agents in targeted therapy [93]. Importantly, the network approach enabled us to predict the effect of PLK1 or Cyclin B1 perturbations on cell viability without having performed any experimental perturbation of these nodes. We tested this prediction by treating the SKMEL-133 cells with the PLK1 inhibitor (BI 2536) and measuring cell viability response with the Resazurin assay (Figure 9). The experimental tests revealed that cell viability IC50 for the PLK1 inhibitor is 5.5nM in SKMEL-133 cells and that approximately 99% of the cells were killed with a 15mM concentration.

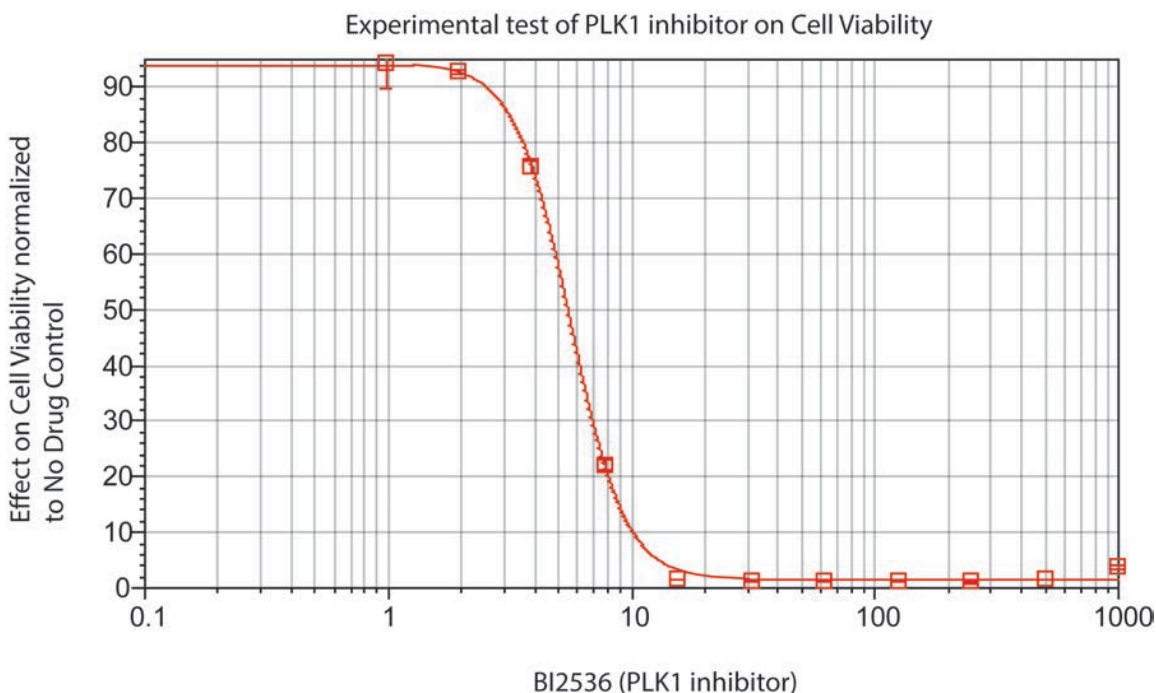

**Figure 9: Experimental testing of computational predictions.** Qualitative analysis of networks in *in silico* simulations nominated PLK1/CyclinB1 as potential targets to kill RAF inhibitor resistant melanoma cells. A validation experiment with the PLK1 inhibitor (BI2536)[23] shows extensive growth inhibition in SK-MEL133 cells (Cell viability IC50= 5.8nm). PLK1 inhibitor is a pure prediction of the approach and was not included in the experimental drug set**.**

The fourth perturbation that yielded a significant change in cell viability was TSC2pT1462 inhibition, which had a central role in the inferred network models particularly in PI3K/AKT pathway. It was regulated by AKT phosphorylation and also interacts with MAPK14pT180, which was upstream of PLK1 and CyclinB1 in the models. Note that SKMEL-133 cell line is PTEN null and constitutively active PI3K/AKT pathway may play a role in drug resistance in this cell line. Thus it is not surprising to us that deactivation of a central player in this pathway leads to reduction in cell viability. The corresponding best drug target is the kinase that leads to phosphorylation of TSC2 on T1462. The effect of each of these four computationally perturbed nodes on cell viability either (i) was consistent with perturbation of immediately adjacent nodes already tested experimentally, or (ii) was a natural consequence of the genetic background of the tested cell line (i.e. PTEN null) or (iii) nominated a new drug target predicted to be highly effective and validated experimentally.

## Discussion

**Beyond classical molecular biology.** Here, we describe a combination of experimental and computational methods, in the field of network pharmacology, to construct quantitative and predictive network models of signaling pathways. The particular contribution is a set of algorithmic advances, which we adapted from statistical physics to infer network models in sizes and complexities not reachable by classical gene-by-gene molecular biology. The necessity of inference of complex network models stems from the fact that classical methods, in which a small number of perturbation experiments lead to functional description of carefully selected sets of genes and gene products is reaching technical limits. High-throughput proteomic and genomic profiling technologies provide much richer and more complex information about cellular responses than can be analyzed by a scientist's thought processes. At these levels of completeness and detail, predicting changes in physiological attributes from molecular data requires computational modeling and quantitative analysis. Our quantitative network models not only capture already known biological interactions but also nominate novel interactions and detect complex regulatory mechanisms, such as multiple feedback loops, in specific biological contexts. The quantitative analysis of molecular and cellular behavior in these models provides detailed understanding of the coupling between signaling processes and global cellular behavior. Such understanding is hard to achieve by reductionist approaches that focus on the relation between single or few molecules and cellular processes. Furthermore, we provide a systems biology platform to predict the cellular (i.e., proteomic and phenotypic) response profiles to multiple perturbations such as those in combinatorial cancer therapies,

**Network pharmacology in the era of personalized medicine**. This investigation constitutes a proof-of-principle of a particular technology: a combination of a network inference algorithm and a technology for perturbing cells and measuring their response. The overall context is network pharmacology, by which we mean the science of using network models to derive and then test effective therapeutic combinations. The particular challenge, in cancer biology, is the complexity and individual variation of genetic and epigenetic alterations that are plausibly cancer-causing and modulate the response to therapy; and, the emergence of resistance to successful targeted therapies, such as EGFR inhibitors in lung cancer or RAF inhibitors in melanoma. In our view, the fairly fragmented gene-by-gene classical methods of molecular biology, while extremely powerful as a reductionist method, are reaching a clear limit. Such methods struggle with effects such as "cross-pathway coupling", "multigenic diseases" or individual variation of response to therapy. More comprehensively quantitative and computationally predictive methods ("systems biology") are likely not only to increase our predictive abilities but also save substantial overall effort by computationally testing large numbers of cellular states on a large variety of genetic backgrounds and in exhaustively explored perturbation conditions.

**Power and limitations of the belief propagation method for network modeling of signaling pathways**. The implementation of BP algorithm enables us to infer much larger network models than were reachable with Monte Carlo search methods. First, we used biologically realistic toy data to illustrate the dramatic speed improvement of the BP method over standard Monte Carlo methods. BP convergence times scale favorably with the size of the models at no measureable loss of accuracy. This is a desirable property for constructing large network models *de novo*. Even in the biologically realistic conditions of noisy data from sparse perturbations conditions, BP inferred most of the dominant interactions in the data-generating network. Though a fraction of the likely BP interactions were false positives, these false positives are consistent with the data and connect structurally correlated nodes. The assumptions inherent in this BP implementation are potential limitations to the overall efficacy and accuracy of our approach. However, assumption 1, while important, does not dramatically affect the results beyond a threshold discretization resolution. Furthermore, since BP performed as well or better than MC, which employed neither assumption 3 nor 4, we suspect by process of elimination that assumption 2 is the largest contributor to the observed limitations of our current method. For real biological data, the quality of BP's performance depends on the applicability of the model equations and the richness and quality of the perturbation data used for training.

We have previously published an MCF7 breast cancer cell line dataset, for which network models were inferred with a nested Monte Carlo search algorithm (CoPIA) [94]. Reassuringly, many of the strongest interactions from CoPIA are recovered by the BP method (Supplementary Text-S2, Figure S10). We have also modeled the same SKMEL-133 dataset with Gaussian Graphical Models, a popular probabilistic model that goes beyond pairwise correlations to distinguish between direct and indirect correlations. A comparison to the BP derived interactions suggests non-trivial overlap in the strongest couplings. A more detailed analysis is available in the supplement (Supplementary Text – S3, Figures S11 and S12).

**Confirmed and predicted interactions in malignant melanoma cells.** We have been able to capture the known interactions in MAPK and AKT/PI3K canonical pathways through de novo network modeling. Moreover, we predicted potential logical interactions such as a bidirectional interaction between RB and MAPK pathways and a potential feedback loop from PI3K and MAPK pathways acting on Src activity through inhibitory phosphorylation at SrcpY527. We determined the signaling events, which influence the cell viability phenotype with in silico perturbations. The quantitative analysis reveals that three signaling events have considerable influence on the cell viability. The in silico perturbations of CyclinB1 and PLK1 lead to comparable reduction in cell viability. As neither CyclinB1/PLK1 nor their direct regulators were inhibited in the perturbation experiments this result is a nontrivial prediction. We also tested and validated this prediction by measuring the cell viability after treating melanoma cells with a potent and selective PLK1 inhibitor. As validated by experimental tests, PLK1 inhibitor reduces cellular growth significantly in melanoma cells even when treated with the drug at nanomolar concentrations (Figure 9). Moreover, none of the *in silico* perturbations of the MAPK pathway lead to a significant change in cell viability, which is consistent with the experimental responses to perturbations acting on MEK and BRAFV600E and RAF inhibitor resistance of SKMEL-133 cells.

**The problem of drug specificity**. We are aware that drugs do not usually have a single specific target. While this is potentially problematic for modeling and simulating drug effects, we find that the inference is driven largely by the correlations in the data arising from the effect of perturbations on the overall system. We estimate that BP is most sensitive to strong and untrue assumptions about the direct effects of a given drug. The use of 'activity nodes' is an indirect way of dealing with drug off-target effects. These nodes, which are perturbed but not measured, represent the coupling of a drug with rest of the system. Such coupling includes both specific and off-target effects. BP may infer interactions from these activity nodes to any other measured node thus representing multiple targets for the drug used. Drug specificity also affects the predictive power of the resulting models. All effects from the AKT inhibitor, for example, are assigned to single knockdown of the aAKT node, even if other targets of the AKT inhibitor are partially responsible for the measured outcomes. Therefore, any simulation-based prediction regarding perturbation of the aAKT node will be tethered to the off-target effects of the AKT inhibitor used in training. The complete solution of the drug specificity problem requires a more comprehensive and systematic analysis to determine the effect of off-target effects on quality and predictive power of inferred models.

**Failed predictions and optimal design of experiment.** The detailed analysis of the predictions reveals few outlier points, where the prediction of the response profiles from incomplete datasets fails. These outliers are in two major categories. In the first category, (mis)predictions arise due to measurements with very low signal to noise ratios and high experimental uncertainties. The mis-predicted S6pS240 levels and cell viability phenotypes in some of the perturbation conditions fall into this category. Note that the models are trained and simulated after logarithmic conversion (i.e. the readouts are first normalized to their counterparts in no-drug conditions and the log-2 of the ratio is computed), which exaggerates the errors for low signals. In linear space (i.e. when all values are converted back to linear space) no such outliers are observed suggesting that those outliers can be considered as an artifact of our analysis in logarithmic space. In the second category, (mis)predictions arise due to insufficient experimental constraints. (Mis)predictions of AKTpS473 levels after mTOR inhibition falls into this category. mTOR inhibition leads to an increase in AKT phosphorylation possibly due to the disruption of a feedback loop. In our current analysis, mTOR inhibition and steady state measurements on AKT phosphorylation are the sole experimental inputs to detect the changes in this feedback mechanism. When the mTOR inhibition is withheld in leave k-out calculations, experimental constraints become insufficient to describe the regulation of AKT phosphorylation leading to the (mis)predictions. A systematic experimental design to enrich the perturbations, better characterization of the AKT phosphorylation dynamic range and richer proteomic measurements on this part of signaling pathways are possible ways to improve the quality of these particular predictions. In general, careful optimization of perturbation conditions (drugs, shRNA) and observations (protein arrays, mass spectrometry target list), within available resources, would significantly enhance the predictive power in this type of model.

**Predicting effective novel drug combinations with BP based network modeling.** Predicting the effect of drug combinations is highly challenging and has been the subject of many studies, e.g., [95,96](Geva-Zatorsky *et al*, 2010 ; Gupta *et al*, 2010). When a large repertoire of targeted drugs are available for screening, the search space for useful combinations of two or more drugs is combinatorially complex. Moreover, therapeutically promising drug combinations are not limited to simultaneously introduced perturbations. Potentially useful drug combinations may consist of combinations of relative doses, for two, three and four drugs applied simultaneously or sequentially after well-defined time intervals. Identification of such more complicated combinations through experimental screening tests is prohibitively cumbersome. Here, we provide a potential solution to this problem in that *in silico* screens using predictive network models can cover a large space of possible drug combinations. The predictive power of the models derived here is apparent from the reasonable accuracy of predicting the results of withheld experiments using a leave-k-out cross validation procedure. Network models inferred from series of perturbation experiments can thus be used to predict the responses to novel combinations, which were not or have not yet been experimentally tested. One can enlarge the search space for drug treatments from a few hundred experimentally screened combinations to tens of thousands of computationally tested combinations and guide subsequent highly efficient top-candidate experimental screens.

Beyond the computational power of well-constrained and robustly derived network models, one may expect to achieve a conceptual understanding of the principles of epistasis of drug effects and the mechanisms of resistance to targeted therapeutics. For example, the initial 'network' response to drug intervention on the scale of minutes to days may be indicative of subsequent epigenetic and genetic changes in a population of treated cells that represent the long-term and hard-to-treat emergence of drug resistance. In this context, reliable dynamic network models may be an excellent guide to strategies for blocking the emergence of resistance in the first place. The pre-clinical consequence is the selection of combinations of therapeutic interventions that not only are effective in slowing the proliferation or promoting the elimination of cancer cells, but also are good candidates in clinical trials aiming to counteract resistance to otherwise effective treatments, such as RAF inhibitors in melanoma or AR inhibitors in advanced prostate cancer.

**Method improvement.** While the approximate solutions to the problem of network inference deliver interpretable biological results, there is much room for improving the power and information value of the method. For example, when time-dependent response measurements after perturbation are available, one can use analogous, extended algorithms to infer probability distributions over (time-independent) interaction parameters $W$ that describe the behavior of a time-dependent system. Another extension of the formalism is the use of more complicated forms of the differential equations, such as those of enzyme kinetics. Also, even in the current approximation, careful design of experiments selecting a minimal set of maximally informative perturbation conditions would increase the efficiency of this experimental-theoretical approach.

A straightforward and powerful extension is the systematic use of prior information, in the form of quantitatively expressed network models adapted from the current scientific literature or pathway databases. Such prior information is easily incorporated as a set of additional constraints on the probability distributions over $W$ in the cost function (Equation 3). We are also actively pursuing a method for the systematic inclusion of prior knowledge interactions from curated databases. On the experimental side, measurement of richer phenotypic attributes of cells, such as apoptosis or cell cycle arrest, as well as markers of differentiation states would much increase the predictive power by providing more links between molecular and phenotypic quantities.

The network pharmacology approach described here provides a strong tool for a system level description of signaling events in cancer cells and moreover, it presents a step forward in quantitative prediction of responses of cancer cells to drug perturbations. Beyond cancer biology, there is no reason to believe that the proposed technology cannot be used to derive accurate quantitative and predictive network models for biological cellular systems in general, provided sufficiently diverse experimental perturbations (such as systematic shRNA) and sufficiently rich readout (such as protein mass spectrometry, metabolic profiling and cellular image analysis) are accessible. In this way, we hope to extend the power of classical molecular biology to a broad spectrum of cellular systems with targeted, and possibly clinical, applications.

## Methods and materials

### Materials

**Choice of drugs and drug concentrations.** Eight small molecule drugs targeting mainly the MAPK or AKT/PI3K pathways were chosen based on the knowledge of target specificity and relevance for exploring BRAF signaling in SKMEL-133 cells (Figure 5). In order to select an appropriate drug concentration for the RPPA assay, Western blots were used to measure the dose-response effect of each drug on its the presumed targets or downstream effectors (Supplementary Figure S4).

**RPPA and Western blots.** For Western blotting and reverse-phase protein arrays (RPPA) assays, BRAFV600E mutant SKMEL-133 cells were grown in 6-well plates to around 40% confluence in RPMI-1640 medium containing 10% fetal bovine serum (FBS). In a series of perturbation experiments, cells were perturbed with 8 drugs either singly or in paired combinations, and harvested after 24 hours by collecting and freezing the cell pellet. Non-perturbed control cells were treated with drug vehicle (DMSO) for 24 hours (elsewhere called "no-drug control"). Cells were thawed, lysed and protein concentrations were determined by the Bradford assay. Protein concentrations were adjusted to 1-1.5 mg/mL and proteins denatured in 2% SDS for 5 minutes at 95°C. For RPPA, cell lysates were spotted on nitrocellulose-coated slides in Gordon Mills' laboratory at MD Anderson Cancer Center, as described previously [81] and stain with antibodies. Each sample was represented in triplicates originating from three different biological samples (wells). The resulting antibody staining intensities are quantified using the MicroVigene automated RPPA module (VigeneTech, Inc.) and normalized as describes in [81].

**Resazurin cell viability assay.** Cells were grown in 6-well plates and perturbed same as those in RPPA assays. After 72 hr drug treatment, resazurin (Sigma-Aldrich, Catalog # R7017) was added at a final concentration of 44 μM to each well and the fluorescent signals were measured after 1 hr incubation, using 530 nm excitation wavelength and 590 nm emission wavelength. For control wells (0hr drug treatment), the fluorescent signals were monitored after 4hr incubation. Standard curves of cell numbers were generated as well to back calculate the cell numbers in different wells.

*Modeling of signaling networks in a melanoma cell line*

**Setting up belief propagation calculations for network inference.** Network models are constructed using the measured proteomic and phenotypic response profiles to drug perturbations as experimental data. The reported network models contain 25 nodes and are based on 44 experimental constraint sets. Each measured protein level is log normalized with respect to its measured level in the corresponding no-drug control condition. For quantification of the activity nodes, see below. The probability distribution for each possible edge strength in the system is computed using the belief propagation algorithm. In the current implementation, the edge strengths can assume values within the interval [-1, 1] with discrete steps of 0.2. The initial messages are sampled uniformly from a random distribution and the BP algorithm is run until the difference between messages in consecutive iterations is less than $10^{-6}$. A systematic approach is taken to ensure the right approximate connectivity in the network models in order to prevent both non-descriptive, sparse models and over-fit, highly connected models. Very sparsely connected network models have discontinuities in information flow and low predictive power. Over-fitted models lack generalizability of predictive power. Empirical representations of signaling networks in the extant literature are fairly sparse, with approximately 1-2 edges per node. We take guidance from these generally intuitive models and aim for an average of about ~1.5 edges/node as desired order of connectivity in the instantiated network solutions derived here. The temperature scaling constant $\beta$ and the complexity penalty term $\lambda$ are the parameters in the BP algorithm that influence average network connectivity. To reach the desired order of connectivity, these two coefficients are adjusted using a series of systematic belief propagation (BP) calculations with $\beta$ and $\lambda$ in the interval (0, 5.0]. In each BP calculation, the most probable value observed in probabilities is assigned as the edge value, for each interaction; this allows the calculation of the total number of edges with non-zero values and thus of the average connectivity. As a result, $\beta$ and $\lambda$ values are chosen such that the desired level of connectivity is ~1.5 edges per node. These values ($\beta=2$, $\lambda=5$) are used for the computation of the instantiated solutions in subsequent decimation calculations. For each node, $\alpha_i$ is taken as 1 and $\varepsilon_i$ is estimated from the dynamic range of each proteomic measurement sampled in the biological dataset. The $\varepsilon_i$ and $\alpha_i$ parameters are further optimized with a gradient descent algorithm.

**Inferring distinct, executable network models of signaling.** Next, distinct model solutions are computed with the BP-guided decimation algorithm. The edge parameters in each model are further optimized using the Pineda/gradient descent algorithm [33], relaxing the assumptions of discrete edge values and factorized probability distributions (Equation 7d). This gradient descent algorithm includes optimization of both $\varepsilon_i$ and $\alpha_i$ parameters. The optimized models are ranked according to their model errors and the best 100 models are used to compute averages of predictions and the graphical representation of the set of best solutions as the average network model (Supplementary Figure S6). The average network model is for summary and illustration purposes and is not in itself executable. Instead, prediction are made by simulation of individual, or sets of, instantiated network models.

**Simulating signaling network models.** Each network solution is simulated individually with specific virtual perturbations according to the model equation 1 until the system reaches its steady state (Supplementary Figure S9) (i.e. until no system variable changes in consecutive steps of simulation within machine precision). The DLSODE integration method (ODEPACK) [97] is used in simulations (default settings with, MF =10, ATOL= 1e-10, RTOL=1e-20) is used for simulations. Trajectories for the best 100 model solutions yield an ensemble of predicted outcomes in response to *in silico* perturbations.

**Activity nodes.** "Activity node" is a technical term defined within the context of applied perturbations and derived network models (Figure 7-Right, Supplemental Figure- S15). Each activity node quantitatively represents a molecular process or reaction, such as phosphorylation, involving a particular protein (or other signaling molecule) that is affected by a perturbation agent. Since we measure protein and phospho-protein levels and do not directly measure the biochemical activity of any kinase, the activity nodes stand in for the effect of each drug perturbation on the biochemical activity of the drug targets. At a basal level (no perturbation), the quantitative measure of an activity node is equal to the activity level in 'no drug' control experiments and is set equal to 0 as a reference point. In the presence of an inhibitor molecule affecting a particular activity node, it is calculated based on the influence of the drug on its presumed immediate or downstream target validated with Western blot experiments (Supplementary Figure S4). We demonstrate this with an example (i.e., quantification of the MEK activity node (aMEK)). We measure the strength of a MEK inhibitor by measuring the phosphorylation of its downstream target MAPK1 at residue T202. If the level of MAPK phosphorylation inhibition is 55% compared to a 'no drug' condition, the strength of the inhibition is $u_i=\log2(0.55)=-0.863$. Based on the model equations (Equation 1), the value for the activity node is $X_a = (\beta/\alpha) \tanh(u_i) = -0.697$. The $\alpha/\beta$ is initially assumed to be 1 and refined by the final gradient descent optimization step. Activity nodes are not allowed to have any upstream regulators except the inhibitor since we do not have any direct measurement of the activity node. All of the activity nodes are quantified using the above procedure and the responses from presumed targets.

**Mathematical description of perturbations**. A constant perturbation ($u_i$) acting on a particular node *i* (Equation 1) has an impact on both the time derivative and the final steady state value ($x_i$) of the perturbed node. The set of interactions and their values ($w_{ij}$) are independent of the perturbation. As modeled by Equation 1, the dynamic properties and steady state value of node $x_i$ are a function of the combination of influence from all upstream nodes ($x_j$) connected to node with nonzero $w_{ij}$ and the strength of the perturbation. The perturbation term in Equation 1 models the effect of targeted interventions such as targeted small molecules. The model equation can also incorporate other perturbation forms such as genetic alterations or RNA interference (RNAi). In case of genetic alterations, the impact can be modeled by fixing $x_i$ to a desired value. For example, fixing a particular $x_i$ value to 0 accounts for a homozygous deletion in the corresponding gene. One may also fix the value of $x_i$ to a positive value to model the impact of amplification of a gene product (e.g., DNA copy number change). RNAi perturbations can be modeled similarly to targeted drug perturbations. The perturbations with varying siRNA concentrations have a direct impact on both time derivative and steady state values of its targets. RNAi perturbations, however, would influence all nodes associated with a particular gene (e.g., total level, level of phosphorylated gene product) since the system is perturbed at the gene expression level.

**Toy Data:** We generated toy data based on toy network models in order to test the performance of BP against a known set of true interactions. The toy models are generated by first fixing a topology positive and negative values, which are then assigned a set of real values by drawing from an even distribution between 0 and a maximum strength of 2. The topologies are designed to represent cascade-like hierarchical networks to include parallel chains, feed-forward and feedback motifs. For the analysis focusing on true interactions, the topology was generated with the web-service Gene Network Generator (GeNGe). At the time of this analysis, popular toy data generators such as GeneNetWeaver focus on scale-free like network topologies that are common in gene regulatory networks, but not typical for signal transduction pathways. Given the network model, the data is generated by simulating the model according to Equation 1 in response to external perturbation, until the system reaches a stable steady state. The steady state values for all nodes are recorded in each perturbation condition. In rare cases, the simulations encountered perpetual oscillations and these results were excluded from the final toy data set. These simulations are purely deterministic as no stochasticity was incorporated into the simulations. Noise was added to the data post-simulation. We chose to simulate the dynamics of the toy networks with Equation 1 so as to remove as a source of error the suitability of Equation 1 as a suitable model of the underlying system.

## Acknowledgements

The authors would like to acknowledge the support of Deb Bemis for her help in organizing the draft of the manuscript. We also thank Sven Nelander, on whose work this research is based, for his comments on the draft of the manuscript and his guidance and perspective along the way.

# Supplementary material to: Perturbation biology- inferring signaling networks in cellular systems.

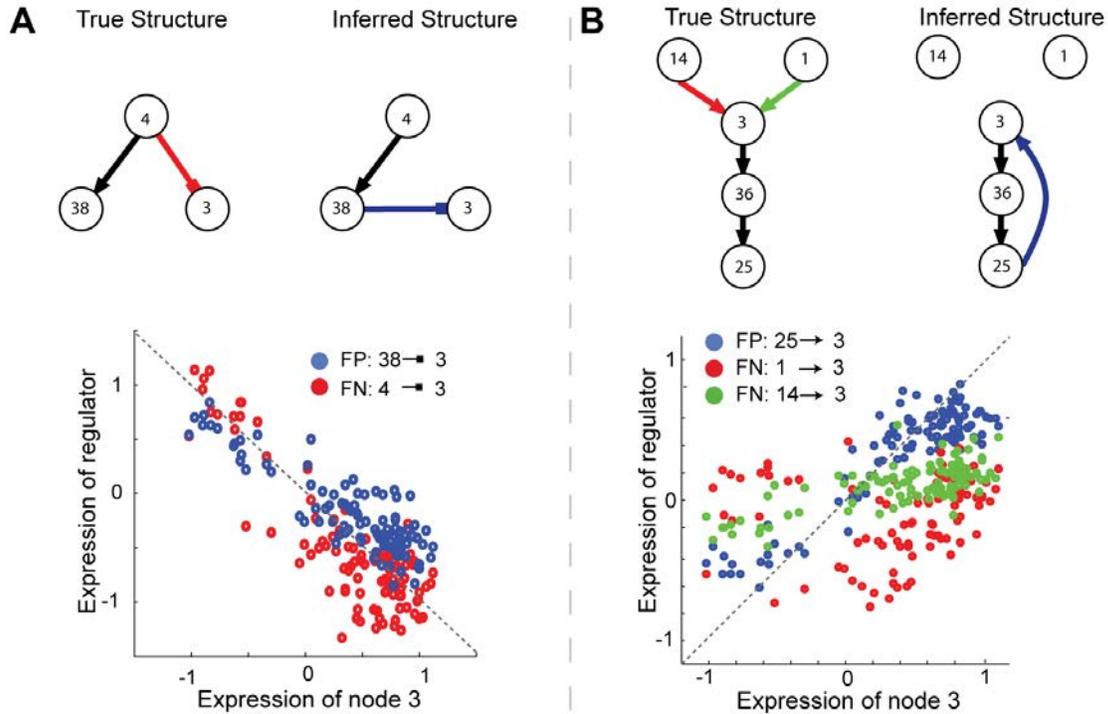

**Figure S1: Comparison of the correlations between false-positives and true-positives.** The nodes connected by false positive edges inferred by BP tend to be highly correlated (blue dots in Figures S1A and S1B), as analyzed here for two example false positives, each from different false positive motifs. Indeed the falsely connected nodes seem to have stronger correlation than the nodes connected by real edges (red and green dots). This analysis indicates that BP is indeed capturing significant correlations in the data although through false interactions. It also indicates that the edge parameters that BP misses (red and green edges) are somehow compensated by the false positive parameter (blue edges). In the absence of temporal data or more perturbations of the nodes in this sub-network, there may not be sufficient evidence to correctly infer the causal relationship between these connected nodes.

**Supplementary Text S1: BP accuracy on better data**

In this short analysis, we wanted to investigate the effect that perturbation strategy has on the performance of the BP data. Currently, we use a set of drugs targeting a subset of all model variables and apply them individually and in all possible pairs to generate the training data (Pair Perturbation Data). We wanted to compare this strategy against so-called higher order perturbations, where many nodes are simultaneously perturbed in a single experiment (Extended Perturbation Data). In each experiment, up to five nodes are randomly selected for perturbation. Furthermore, the set of targetable nodes are restricted to the same set as for the pair perturbation data. This study is performed entirely on a single toy model (the same 40 node network as that analyzed in the main manuscript), simulated to steady state for each perturbation condition described above. The steady state responses were subsequently used to infer the interactions of the underlying data-generating network.

The training patterns that emerged from the pair perturbation strategy tend to be highly correlated, such that most pair combinations provided little unique information. Conversely, the extended perturbation data set produces less correlated training patterns (Figure S2E,). It is clear that the training patterns with higher median correlation would have less information than the training patterns with lower correlation.

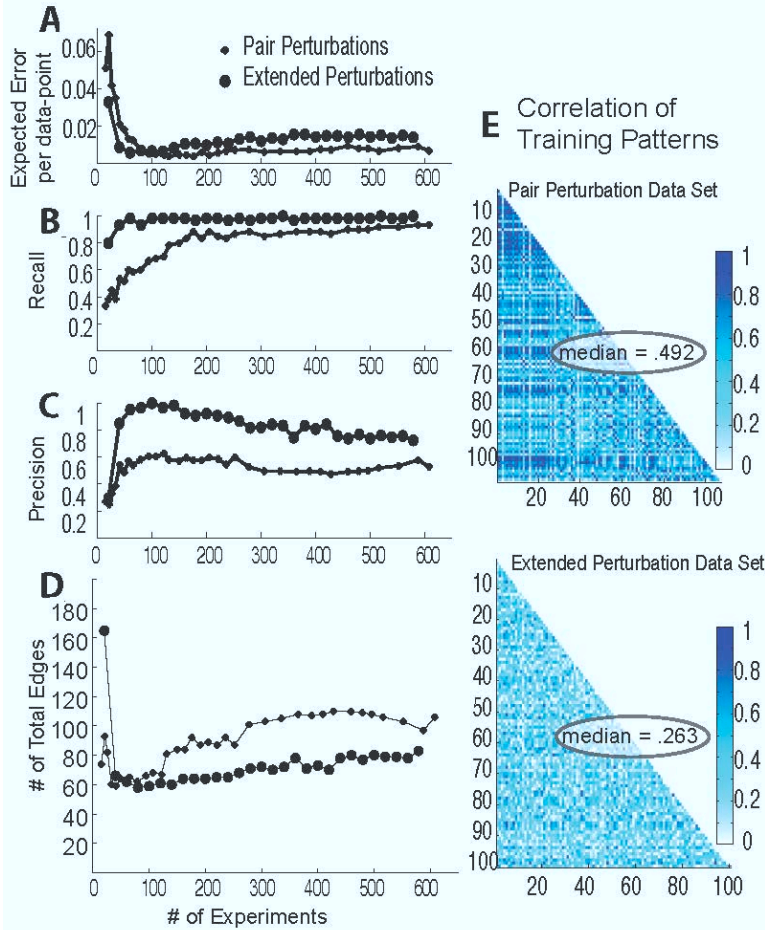

**Figure S2: Alternative higher-order perturbations produce better data for inference with BP than the systematic pair perturbation strategy.** Performance for both datasets is analyzed with mean squared errors (A), recall (B), precision (C), and the total number of BP-inferred interactions. The correlations between the training patterns are shown in analyzed in (E) for both perturbation strategies.

We subsequently evaluated recall and precision (Figure S2B,C respectively) of the BP models trained to the two data sets. The BP models from the extended data set reach a higher recall and do so faster than for the pair perturbation data. Similarly, the precision curve for the extended data set lies consistently much higher than for the pair perturbation set. Finally, as we increase the number of experiments from the pair perturbation data set (by allowing additional nodes to be targeted) we definitely see that we get new information about the true interactions since the recall curve goes up. However, the novel information per additional experiment is fairly low and yields networks with increasing numbers of interactions. These three results confirm that using BP against more informative data produces superior network model, and that different strategies of perturbing the system yield training patterns with different information.

Although we see similar correlations between pair perturbation datasets in biological experiments (data not shown), it is not clear if the alternative strategy proposed here that works on the synthetic data generator would hold in the biological setting due to unexpected drug-drug interactions.

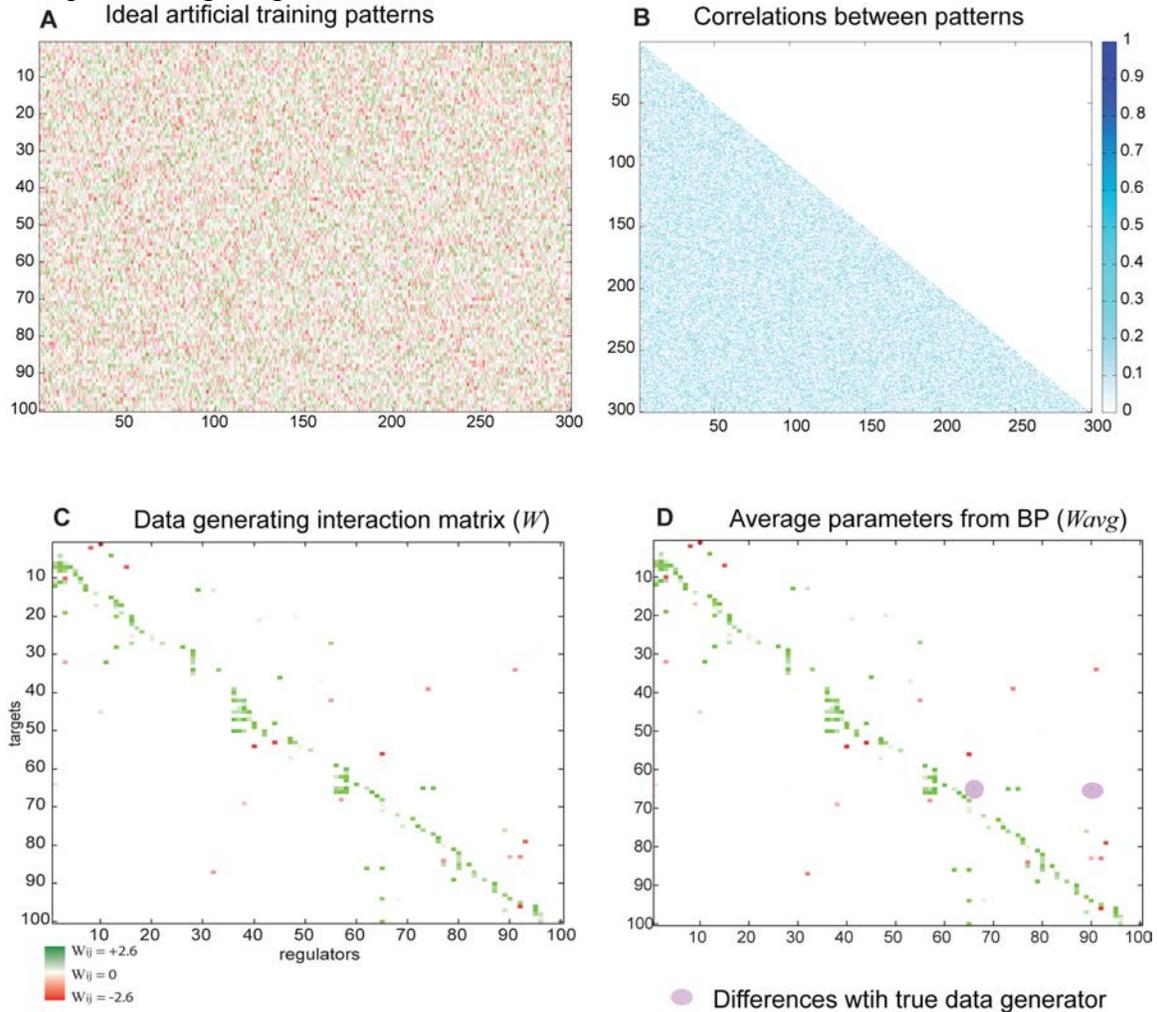

**Figure S3: Perfect BP inference with ideal data**. Totally randomized data (A) produces low correlations between training patterns (B). With this maximally informative, BP reproduces all of the true interactions (D), except 2 with zero false positives when compared against the underlying data generating network (C).

Furthermore, we were interested in looking at BP results when Assumption 2 is not at play. There are many assumptions in the BP method described in the main manuscript. One of the most dangerous assumptions is Assumption 2, which decouples the likelihood function from the dynamics of the model parameters. The synthetic data sets analyzed in the manuscript were generated without assumption 2. In those tests, we see good but imperfect inference of the underlying interactions. To isolate the consequences of assumption 2 and assess the performance of BP when assumption 2 is exact, we have generated this artificial data set (Figure S3A), which is not based on dynamic simulation of the generator network. Instead, we generated 300 training patterns for each model node separately as in Equation 5. We use random, uncorrelated values for the values on the right hand side of Equation 5 to generate values on the left hand side. The result is a set of low training patterns with low correlation (Figure S3B). Inferring network parameters with BP from this data produces near perfect inference with zero false positive and only 2 false negatives (Figure S3D). Although this type of data is infeasible in any biological system, it demonstrates that BP works almost perfect in the extreme case of ample, ideal data. This result also reinforces our suspicion that the effect of Assumption 2 is most likely responsible for the observed limits in BP performance.

## MEK inhibitor: PD0325901

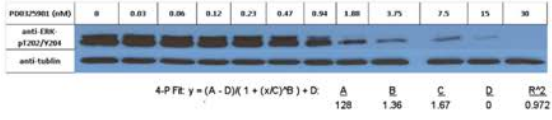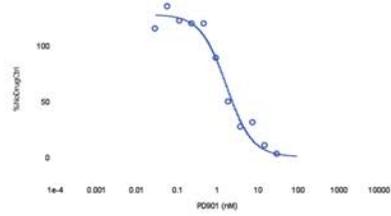

## AKT inhibitor: AKTi-1/2

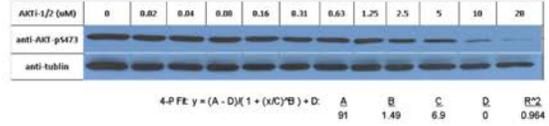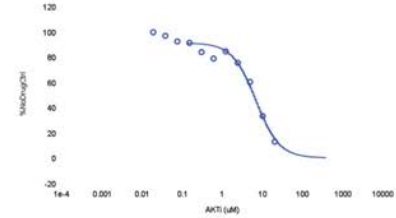

## PI3K inhibitor: ZSTK474

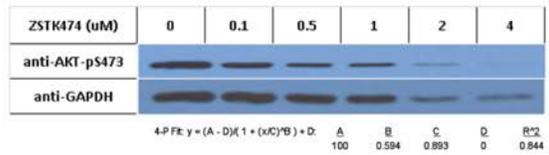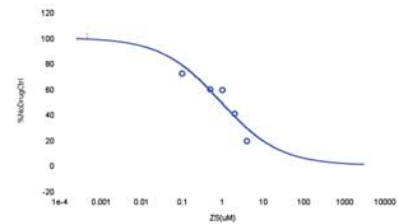

## HDAC inhibitor: HNHA

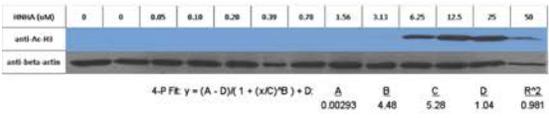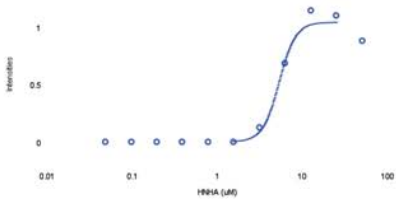

## STAT3 inhibitor: Stattic

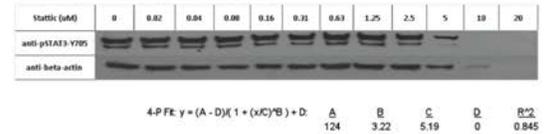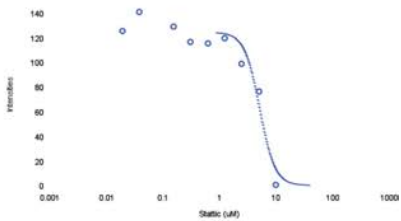

## mTOR inhibitor: Temsirolimus

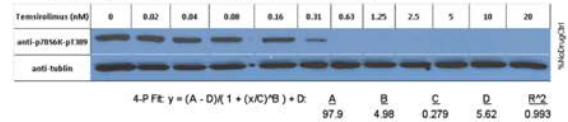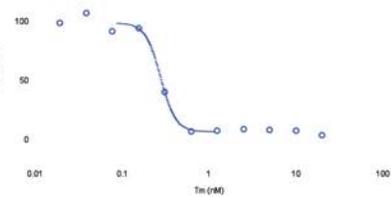

## BRAF inhibitor: PLX4720

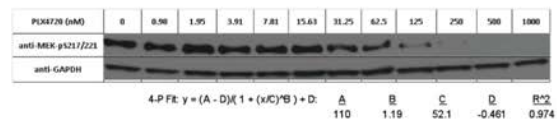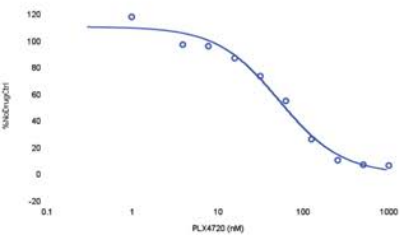

**Figure-S4: Dose response curve of singe agent drug perturbations.**
Dose response curves determined based on Western Blot experiments, shown on top. See table S1 for protein IC40 values, which are calculated with these curves and used in all paired perturbation experiments . Cell viability curve (not shown) was used to estimate PKC inhibitor (Ro-31-7549) IC40.

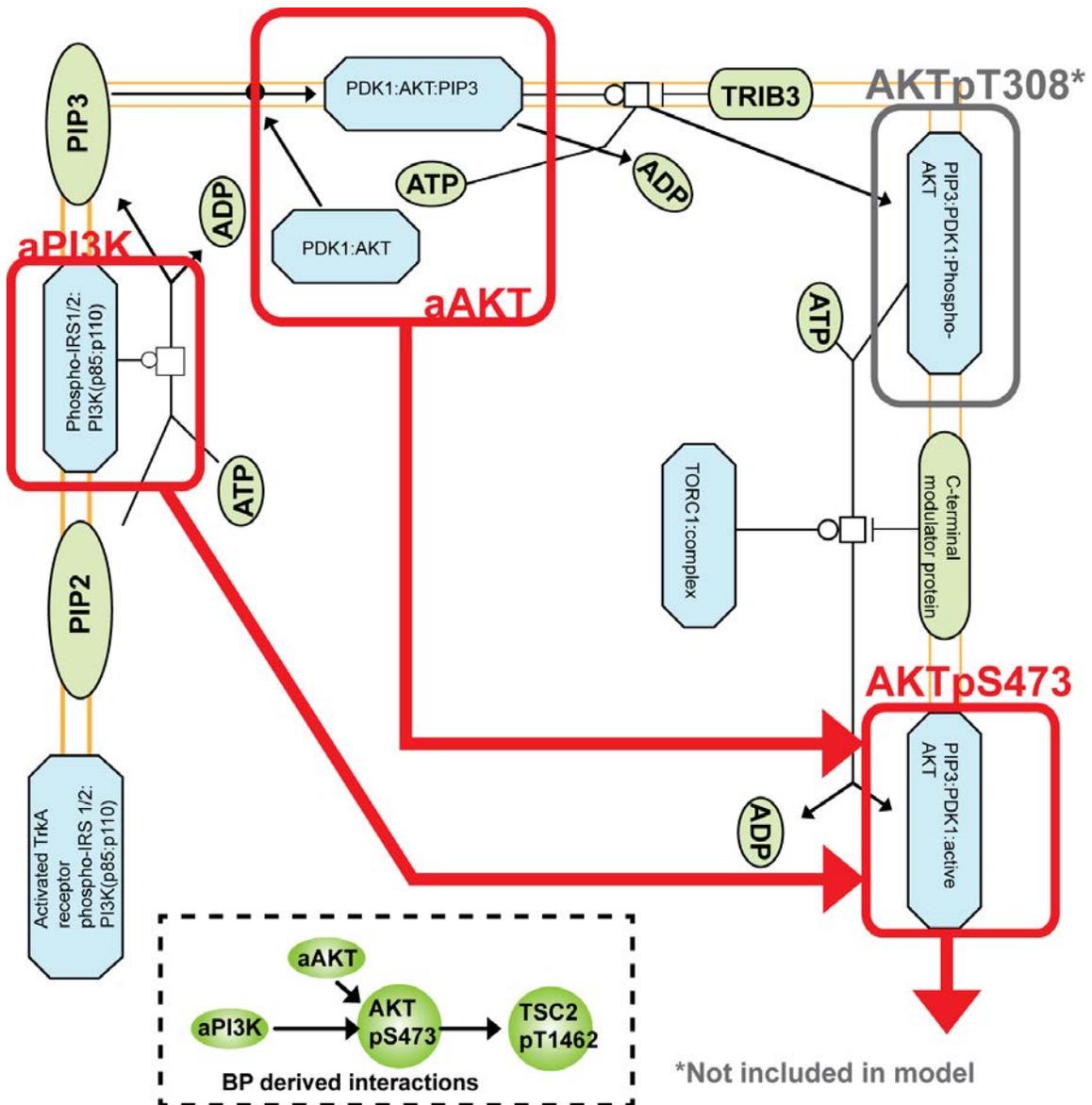

**Figure - S5: Model abstraction in network models of signaling.** A relatively detailed model of biochemical events in PI3K/AKT pathway (i.e. consecutive phosphorylation of AKT at T308 and S473) (reactome.org) and the coarse-grained network model abstraction used in this paper (Red arrows and rectangles). In the model abstraction, PI3K and AKT activity nodes influence the final output of the activation mechanism (i.e. AKTpS473). In princible, we can improve and extend the model with additional measurements such as measurements on PDK1, AKTp308 (circled in gray) etc without altering the model abstraction. Such model improvement requires richer measurements and will lead to a shift from logical to potential physical interactions. Lower left. The PI3K/AKT interactions in the average network model computed with BP-guided decimation.

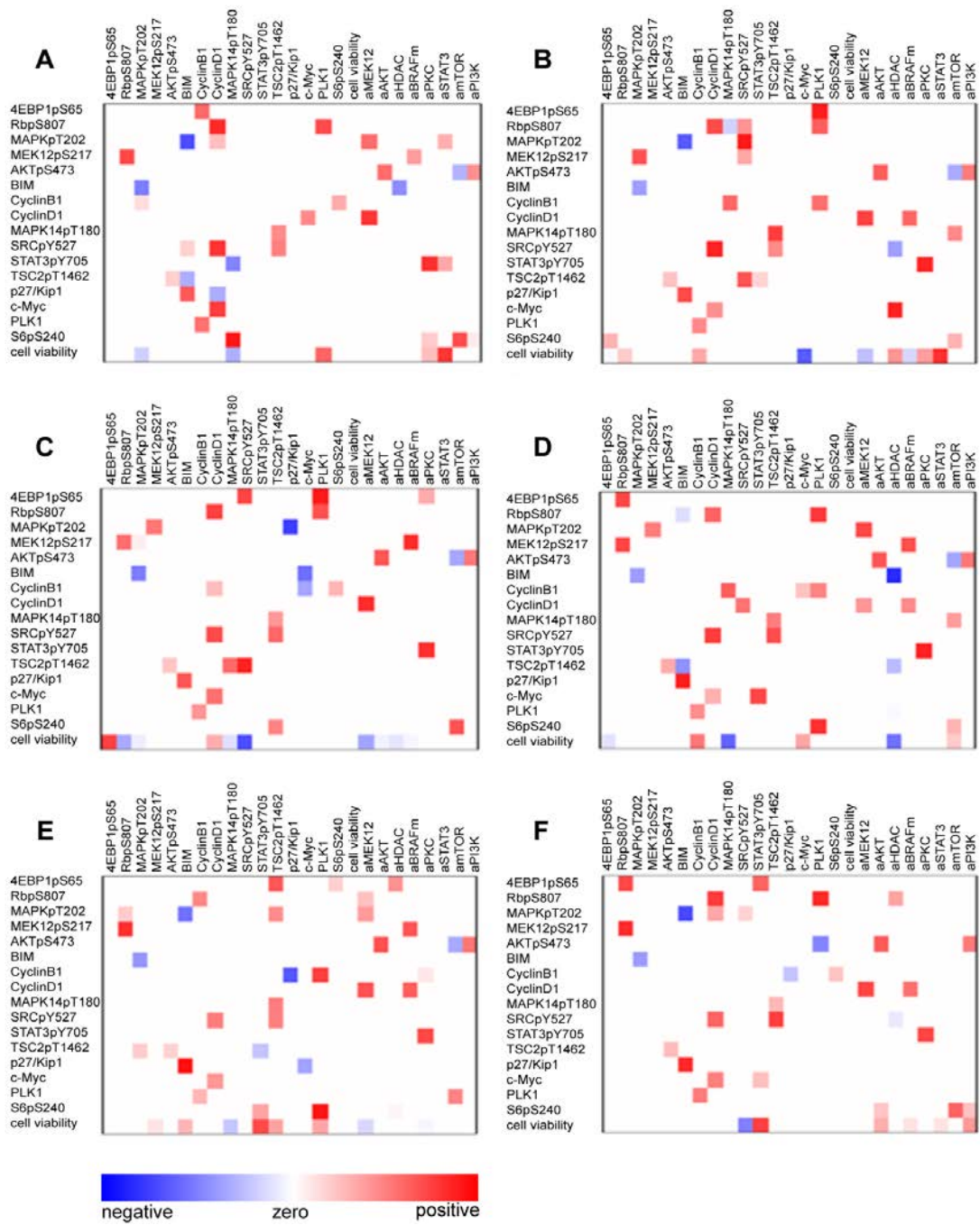

**Figure - S6: 6 of the 10 best network models.** Better models have lower cost. Two nodes (horizontal and vertical list) can be connected by a positive (red) or negative (blue) interaction $W_{ij}$. Although each model is different in detail, each model represents the data well in the predictive sense. The differences reflect genuine uncertainties of model inference, normally not represented in molecular biology cartoon models.

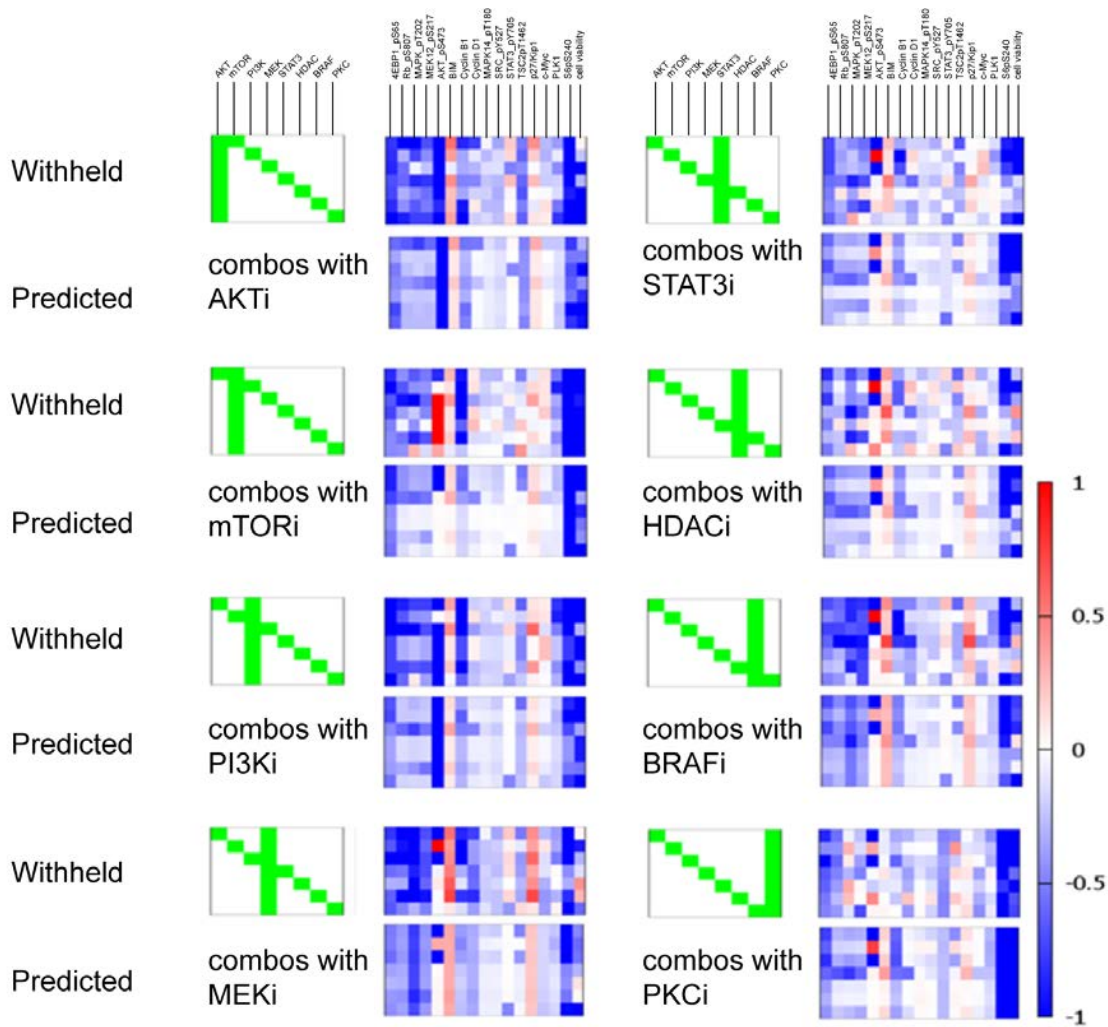

**Figure - S7: Predictive power tested by prediction of withheld data.** The 8 distinct heatmaps of withheld (upper right) and predicted (lower right) respone profiles to withheld perturbations (upper left). For quantitative analysis of predictive power, please see Figure 7 in the main manuscript.

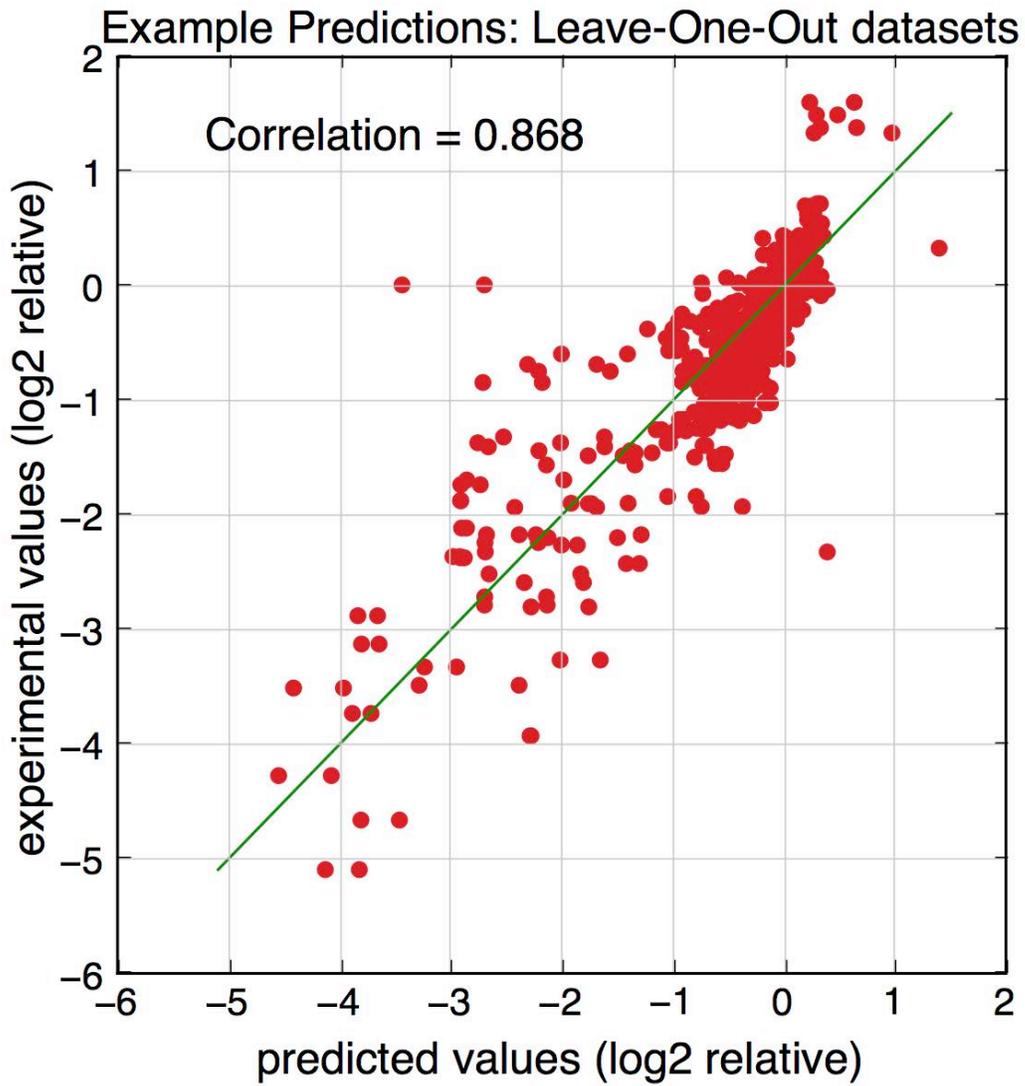

**Figure - S8: Cumulative predictive power.** The cumulative scatter plot that combines the information in Figure S7 (C=0.87).

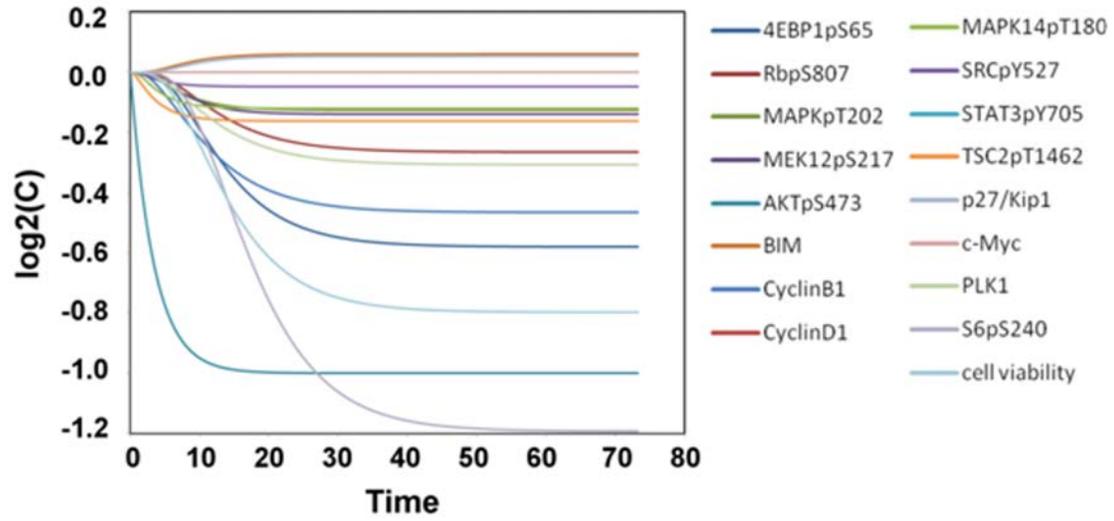

**Figure - S9: Sample simulated trajectory.** Pseudo-time trajectory of a single network model when perturbed with an in silico perturbation on AKTpS473 at virtual IC50 .(i.e. $\log_2([AKTpS47]_{final}/([AKTpS47]_{initial})=-1)$

**Supplementary Text S2: Results on MCF7 data set**

We previously published a different inference algorithm, called CoPIA [94], which was applied to a small data set collected on mcf7 breast cancer cell linesl. The data set consists of 9 measured protein and phenotype responses to
21 perturbation experiments involving 7 drugs. In brief, CoPIA consists of a nested search, in which the outer loop searches over network topologies, while the inner loop optimizes the parameters of that topology with a gradient descent algorithm. Both CoPIA and BP are based on the same model equation as defined in Equation 1. The likelihood function employed in the CoPIA method is a function of explicit numerical simulation, while the likelihood function in BP is based on an algebraic approximation of the steady state, as discussed in Assumption 2. In this short study, we applied our BP based pipeline on the same mcf7 data and compare the results. In this analysis, we used only a single beta (inverse-temperature) and lambda (sparsity) penalty. The BP results are based on the top 100 models from a set of 10,000 models generated from the BP-calculated marginal distributions. The CoPIA results were taken from the edges reported in Table 1 in [94].

BP returns many of the significant interactions determined by CoPIA (Figure S10A), where the edge widths are proportional to the estimated posterior probability from the top 100 CoPIA models. Of all 23 interactions inferred by CoPIA 13 interactions are observed in at least 2 of the top 100 models sampled from BP-calculated marginal probability distributions. Five of the ten excluded interactions were not eligible for inference with BP (blue edges), since BP considers neither self-interactions nor edges directed into 'activity nodes'. Thus only 5 interactions, only 2 of which were high probability edges in CoPIA, are excluded by BP.

BP infers 27 parameters (Figure S10B) with a probability of greater than 20%, of which 8 are also interactions in the top 100 CoPIA models (green). Edge width in the bottom figure is proportional to their frequency in the top 100 BP models. Of particular interest is the large number of high confidence incoming edges into MEK and pERK, both of which were inferred to have self-feedback interactions in the CoPIA models (11 between those two). It is also interesting that with the exception of an inhibitory interaction between pERK and MEK, there are no high confident interactions originating from either pERK or MEK. We do not yet understand why this happens, but we suspect there is a pathological connection between their high connectivity in BP and the inferred self-connectivity in the CoPIA models. Finally, the top 100 BP derived models have a mean squared error of roughly 0.1, which is as low as we've seen.

There is significant agreement among a subset of high confidence edges. This suggests that both BP and CoPIA methods have explored similar areas of the total solution space. However, BP does not return the exact same models as CoPIA. There are many potential reasons why the results are different. The first possibility is that BP's restriction to exclude self-interaction and edges into activity nodes forces BP to search in an alternative region of solutions space. Secondly, CoPIA is based on a fairly simple MC topology search, which is unlikely to explore much of the full solution space. Consequently, the CoPIA results might well reflect a local-minima in the solution space. Regardless, the low mean-squared-error of the top 100 BP models is a positive result.

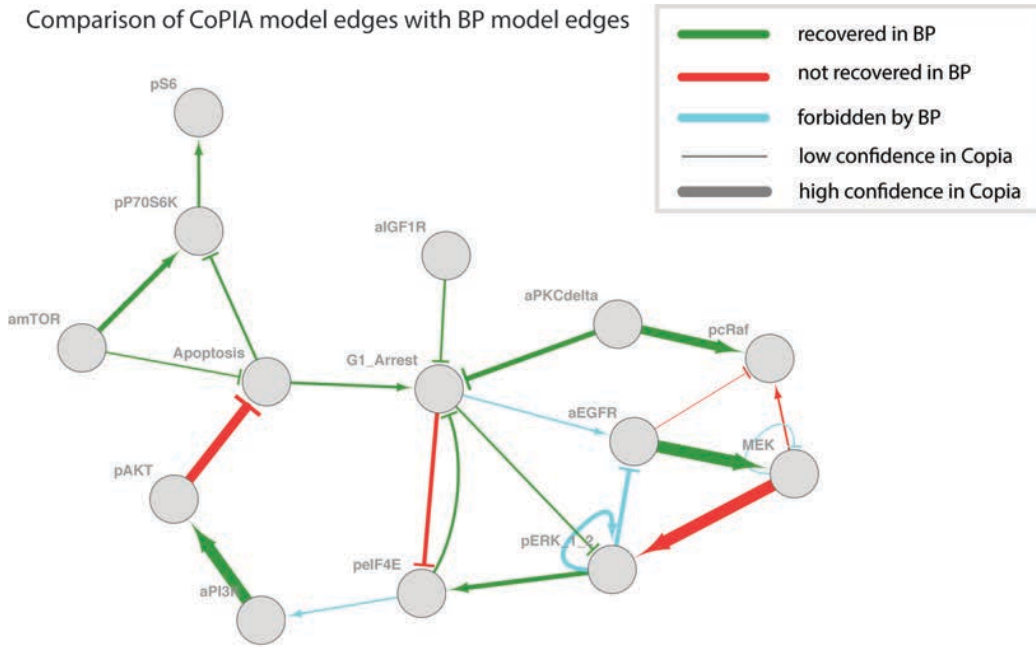

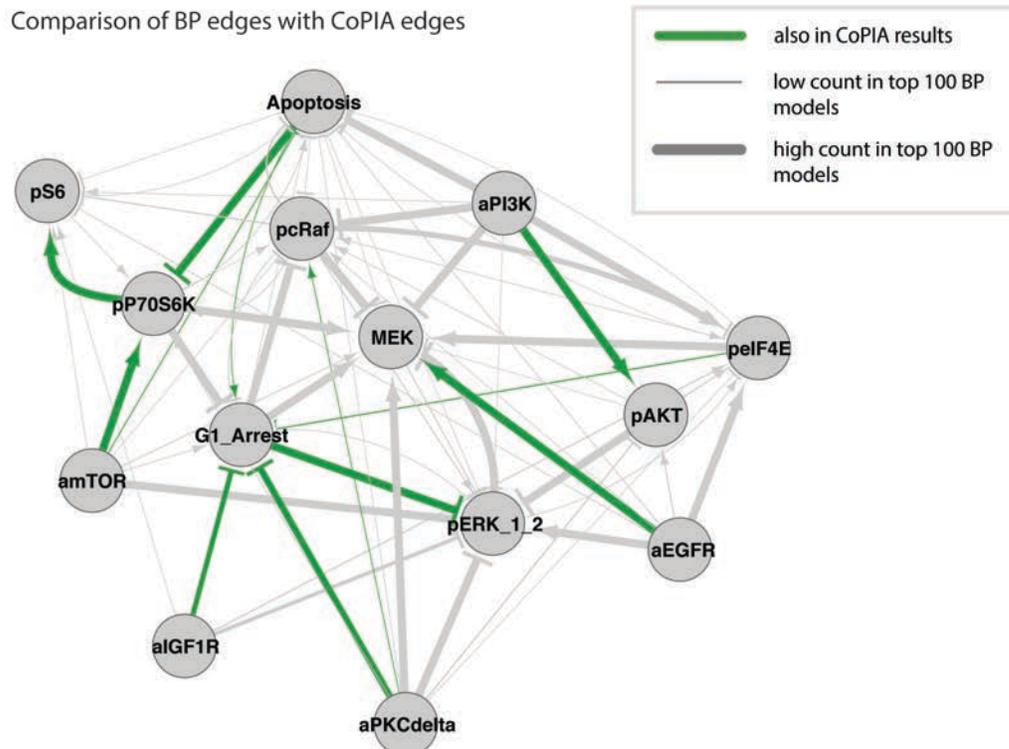

**Figure – S10: BP comparison with CoPIA results on mcf7 perturbation data**. The CoPIA network inference method is an older method based on a nested Monte Carlo

search, and was previously applied to infer network models of mcf7 breast cancer data. The CoPIA network edges are color-coded based on the existence of those edges in BP inference (A) has many agreeing edges (green). Furthermore, most of the significant BP edges (B) are also significant in the CoPIA models (green).

**Supplementary Text S3: Comparison with Gaussian Graphical Models**

Gaussian Graphical Modeling (GGM) is a well-developed model for describing couplings between random variables [98,99]Correlation based networks are typically found to poorly distinguish between direct and indirect associations. Gaussian models are theoretically based on conditional dependencies in multivariate Gaussian distributions, and are likely to better distinguish between direct and indirect associations. Like correlations, couplings in GGMs are symmetric between model variables, and thus the couplings do not distinguish direction. We have adapted standard methods of calculating GGMs, based on inverting the covariance matrix, to infer couplings not only between model variables, but also between drugs and their targets (presuming the targets are included in the set of model variables), denoted *J* and *H* respectively. We use a standard maximum-likelihood estimator to calculate both J and H simultaneously. The goal of this brief study was to compare the strongest couplings from GGMs against those inferred by BP.

The results are summarized in this figure. The GGM calculation is dependent on a single parameter, referred to as the *weight cutoff*, which reweights the training patterns (in this case, perturbation experiments). The cutoff is constrained to be less than or equal to 1, and a cutoff of 1 corresponds to zero reweighting of the patterns. We find that a cutoff of 1 results in the highest likelihood model, given the data (Figure S11A) and the J and H matrices analyzed here are the result of this cutoff. The distributions for both J and H coupling strengths (Figure S11B and S11C, respectively) have the majority of strengths on or near zero, which indicates strong discerning power between strong and weak couplings.

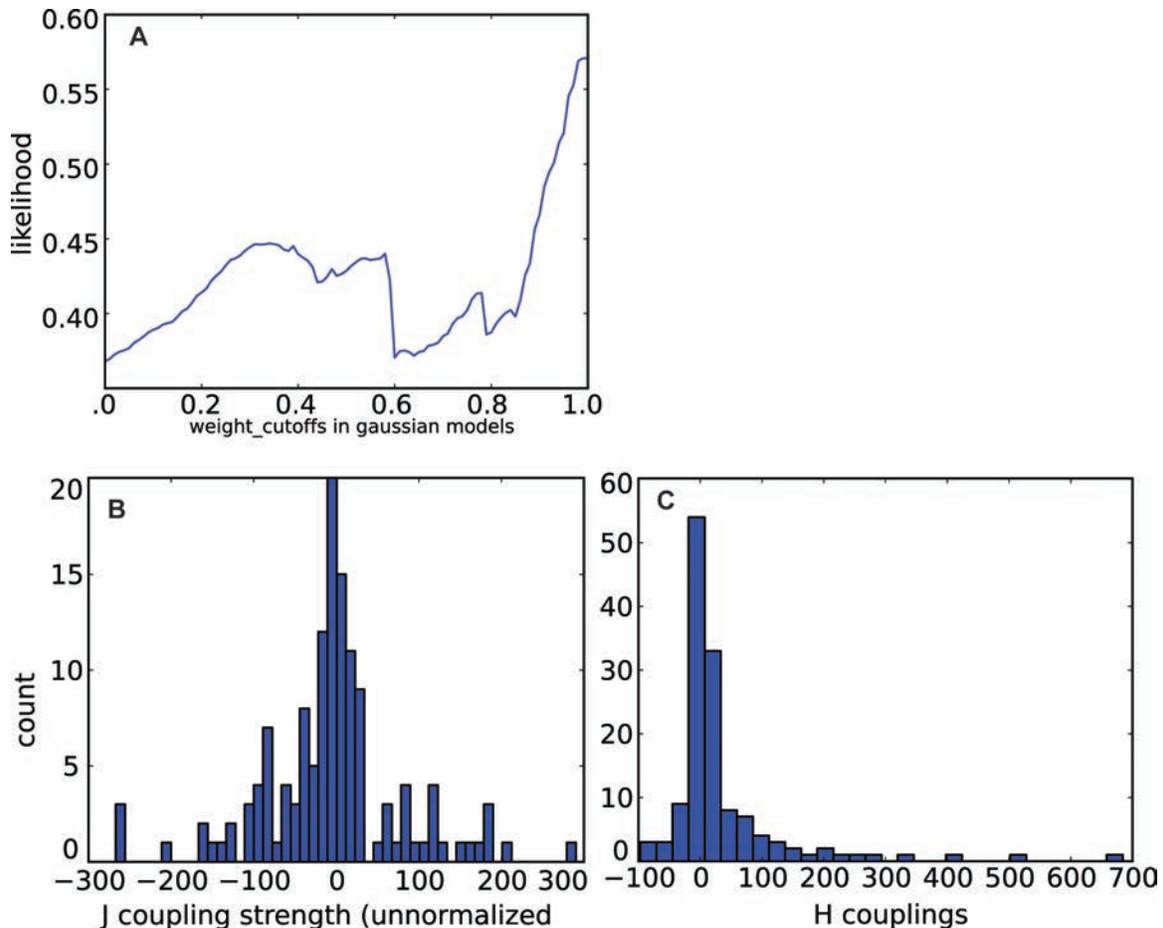

**Figure - S11 : Gaussian model couplings for the most likely cutoff**. The inference with no reweighting produces the highest likelihood estimate (A). The distribution of J and H couplings (B and C, respectively) are strongly centered at zero, demonstrating some discerning power between all possible couplings.

To assess the extent of agreement between GGMs and the BP interactions we must decide where to fix the threshold, below which all J and H couplings are set to zero and thus ignored as edges in a network. If the threshold is too high, we end up with a very sparse network with only the strongest couplings. If the threshold is too low, all variables are coupled and we lose any ability to discern between strong and weak couplings. The agreement with varying threshold is summarized with the Receiver Operator Characteristic (ROC) curve in Figure S12 (top). Crucially, the curve lies firmly above the diagonal line. Two points on this curve that correspond to the maximum F1 score and Matthews Correlation Coefficient (MCC), are independent metrics of the balance between accuracy and precision and are marked on the ROC curve (red and green dots, respectively). Of particular interest is the threshold corresponding to the F1 max (J cutoff of 30), which appears to capture almost 70% of the BP interactions, with only a 40% false positive rate. The network of couplings above the F1-max threshold (Figure S12 bottom left) is still highly coupled with 122 couplings out of a possible 289, of which 18 are captured in the BP network (Figure S12 bottom right). None of the analyses presented here argue in favor of either methodology. The best strategy would depend on the method that can produce the best models in terms of reproducing response to trained perturbations and predicting response to new perturbations, none of which is evaluated here. Instead, we only have evidence that BP and GGMs are non-trivially capturing similar couplings between system variables.

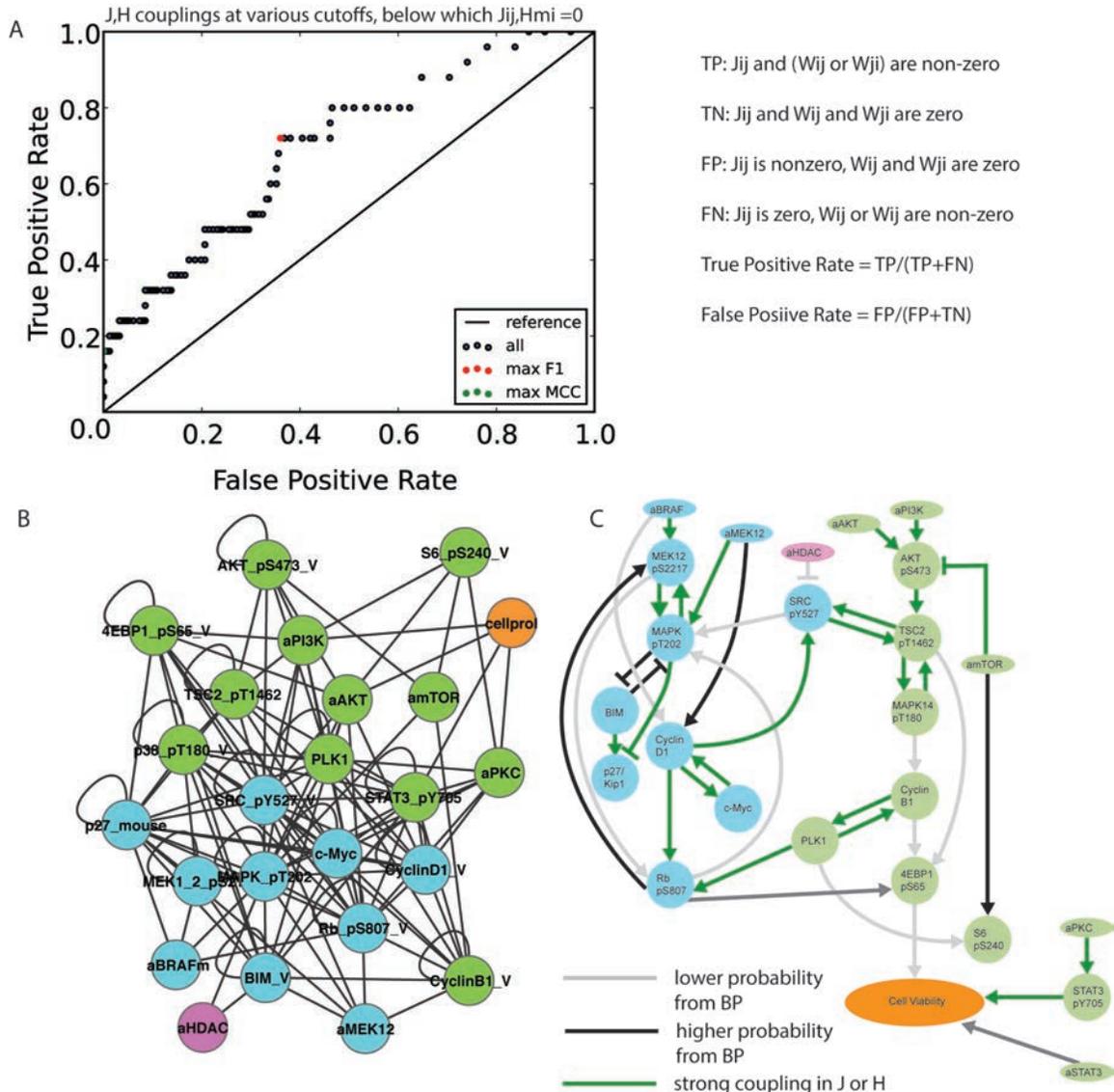

**Figure S12: Comparison of the most likely Gaussian model couplings and those from BP.** The ROC curve (A) lies firmly above the diagonal line. The set of GGM edges taken from the F1-max cutoff (B) has many agreeing edges with those returned from BP (C).

It is important to realize that GGMs, however useful, are not sufficient for our modeling needs by themselves. For instance, they do not infer direction. Additionally, GGMs have no dynamic engine with which to run simulations of the models to predict response to perturbation. Despite these limitations, GGMs remain a well-regarded and mathematically grounded analysis of significant correlations. We are currently exploring ways of incorporating this GGM analysis into the BP-based strategy, potentially using GGMs to define reasonable initial conditions for all marginal distributions.

**Supplementary Text S4: Effects of the discretization assumption (Assumption 1)**

The result of Assumption 1 in the main text is that BP can only approximate an otherwise continuous distribution (the marginal probability of any single parameter) with a discrete distribution. This begs the question of how to pick the scope (maximum weight magnitude) and the resolution or the number of intermediate values (number of bins). In the manuscript we state that both scope and resolution were picked based on heuristic evidence from extensive numerical studies, from which we conclude that 11 discrete values between -1 and 1 are sufficient to guarantee convergence and recover non-trivial variance in the resulting distributions, which balance exploration of a large solution space against the desire for sufficient constraints on the solution space. In this brief, preliminary study, we wish to share our current progress on investigating this issue more quantitatively. This analysis was done on a single dataset from mcf7 breast cancer cell lines.

The choice of discretization affects three main features of the BP algorithm: convergence, consistency and sensitivity. Convergence is the ability of the BP algorithm to terminate to a stable set of marginal probability distributions. Consistency is the agreement between replicate runs of BP on the same data and depends highly on the initial conditions (initial guess distributions), the order of the updates and the discretization. We have recently discovered that a large number of replicate runs of BP sometimes converge to a much smaller number of unique solutions, each of which is a stable local minima in the error landscape. This is a newly discovered and complex issue and is beyond the scope of this analysis.

Sensitivity concerns the ability to capture probability mass that is concentrated between any two discrete parameter values. Consider the hypothetical example where BP is restricted to consider only parameter values -1, 0, and 1 but where the true weight value is -0.5. This means that the update inside BP can only consider all-on (positive or negative) or all-off parameters. The worry is that the entire marginal might also converge to all-on or all-off distributions, such that one value is given 100% probability and all other values have a probability of 0%. Equally problematic would be the result where the probability distribution splits the probability evenly across -1 and 0, such that half of the samples drawn from this distribution would not have a true interaction and even then would be fixed to a value far from its true value.

To investigate this sensitivity, we track the entropy of the set of marginal probability distributions resulting from different discretization strategies. Roughly, the entropy quantifies the degree of uncertainty captured in a probability distribution. Importantly, it does not say anything about the accuracy of the resulting distribution, which is a separate issue. Distributions with absolute certainty have zero entropy, while maximum entropy describes even distributions where all weight values are considered equally likely. We consider three different scopes with maximum weight magnitudes of 1, 2 and 3. For each scope, we consider 8 different resolutions, where the range is divided into equally spaced bins. We consider a minimum of 3 bins covering one negative, one positive and one zero-valued weight.

The metric of interest is the sum of the entropies of all BP-calculated probability distributions. We calculate the entropy for each set of marginal distributions from 100 independent runs of BP to get a sense of the variability of entropies across replicates. One consideration is the choice of normalization, since the maximum entropy (entropy of even distributions) increases with the number of bins. One choice is to normalize the entropy by the maximum entropy for that bin number (Figure S13). Yet another choice is to collapse a distribution into a ternary distribution of negative, positive and zero values, by calculating the aggregate probability mass over all negative, positive and zero values, respectively (Figure S14). These collapsed ternary distributions are fundamentally different from the BP-calculated distributions over only three values in that prior to collapse, the BP considered a larger number of possible values and could capture intermediate positive and negative values.

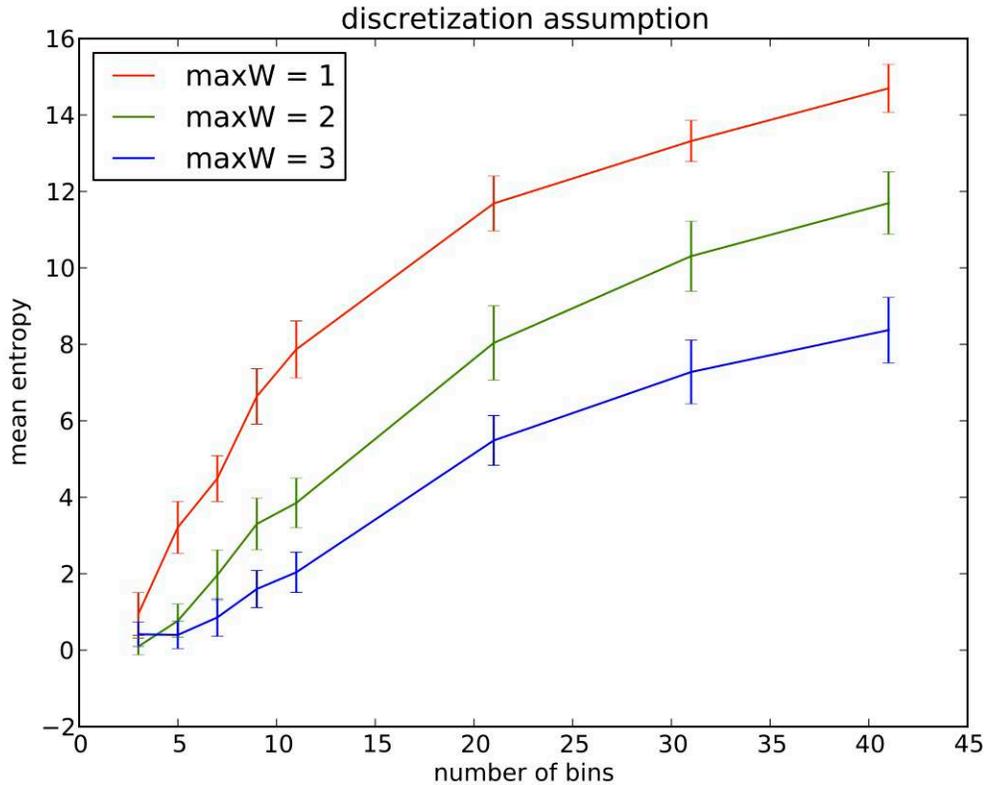

**Figure S13: Entropy as a function of the number of bins**. Entropy is normalized by the ratio of total entropy to maximum possible entropy. This analysis is focused on exploring the effects of different discretization strategies (Assumption 1) on the BP inference.

In both Figures S13 and S14, discretizing over 3 bins produces distributions with low entropies near zero, confirming the suspicion that such discretization tends to yield all-or-nothing marginal probability distributions. These distributions are dangerous since they likely over constrain the search space. Furthermore, true non-zero values are given zero probability and thus would never appear in models drawn from these distributions. In short, appropriate uncertainty is not capture in these distributions.

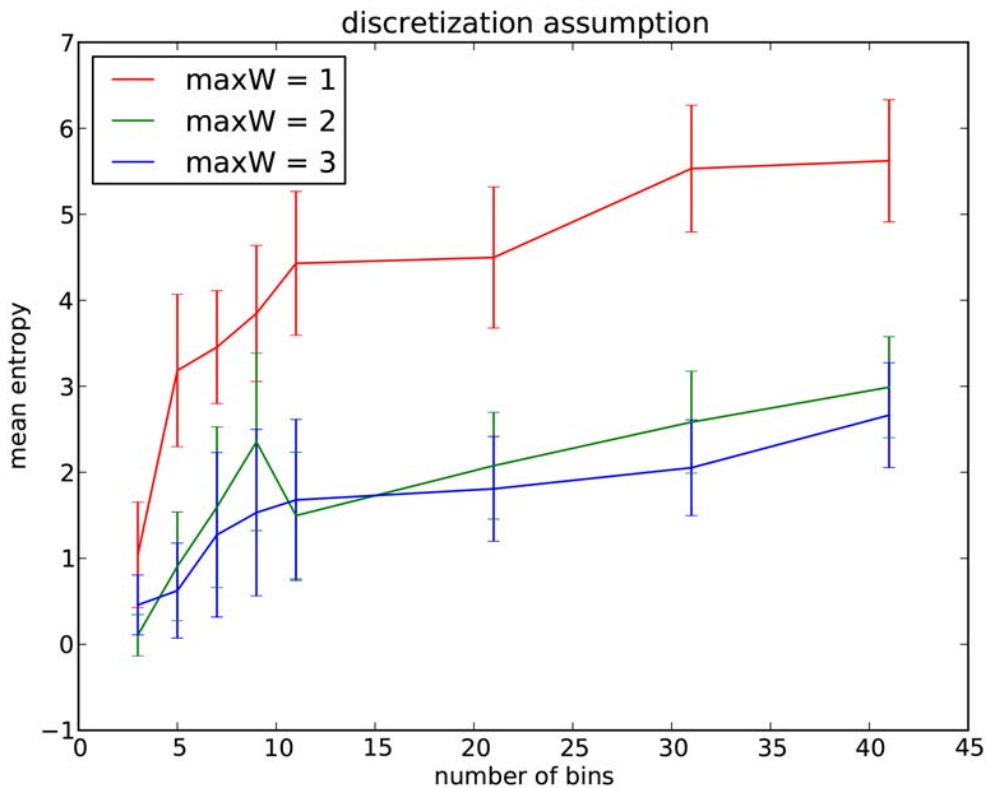

**Figure S14: Entropy as a function of the number of bins**. This time, entropy is calculated on a collapsed distribution over three regions; aggregate probability mass for negative values, zero values and positive values.

Both normalization strategies show increase in the entropies with increasing the number of bins. Both curves are sub-linear and appear to saturate at increasing bins. There are two major differences in the features of these normalizations. First, is that the entropies over the ternary distributions saturate faster and most of the increase in this entropy happens between 3 and 11 bins. Conversely, the entropy increases substantially all the way up to 41 bins in Figure S13. Without supplementing this analysis with comparisons of marginal quality (in terms of accuracy or predictive power), it is unclear which discretization strategy is optimal and whether there is any useful advantage to higher resolutions beyond 11 bins.

The second major difference between Figures S13 and S14 are the vertical shifts in the curves between different scopes. This phenomenon in S13 is easily explained by the fact that increasing the scope for a constant bin number results in a proportional increase in distance between discrete weight values. Assuming that two distributions have identical means and variances, then widening the scope without a changing the number of bins, would concentrate the probability mass over fewer bins resulting in a lower entropy. This is likely why we see uniform decrease in entropies in Figure S13 for increasing scopes. This artifact is not applicable for the entropy calculation in S14. In Figure S14 we notice the vertical shift from high entropies with a scope of 1 to lower entropy curves that are very similar for scope of 2 and 3. The reason for this shift is not yet solved. One hypothesis is that there is more uncertainty when the scope is insufficiently small. If the scope were too small, then the largest magnitude value is insufficient for some parameters although we hope that those values be given a non-zero probability. Because of this inefficiency, BP must explore alternative parameters to compensate for the unexplained variance between observed and expected. This hypothesis also explains why the entropy curves are similar for both scopes of 2 and 3. If the true scope were between 1 and 2, then any scope above 2 would cover superfluous parameter values. And since the entropy in Figure S14 is only a function of the aggregate probabilities of all negative, positive and zero-valued weights, then the entropy is agnostic to the distance between weight values, unlike the entropies in Figure S13.

These results are inconclusive and communicate only our progress towards achieving an objective strategy for picking an optimal scope and discretization that maximizes convergence rates and is sufficiently sensitive to balance the desire to explore parameter space against the attempt to restrict the parameter space to the most likely parameters.

**Supplementary Text S5: The Monte Carlo Algorithm employed in this manuscript**.

We compare Belief Propagation (BP) against a standard Monte Carlo (MC) algorithm, as MC is ubiquitous and well understood throughout the literature. MC also has its roots in statistical physics. If MC is run through enough iterations, it is guaranteed to ultimately traverse the entire solution space and is in this, respect, a greed search algorithm. Such exhaustive searches are prohibitively long and thus an adequate termination criterion determines the run time of the MC. Marginal probability distributions are drawn from MC solutions by summing the frequency each parameter appears and weighting it by its likelihood. BP is guaranteed to converge to the exact marginal in tree-like relationships (no loops) between variables and constraints. However, since our problem is fully-connected, each variable is connected to all constraints, BP is not guaranteed to converge to the true marginal and is only an approximation. Since BP is a method to approximate the same marginal probabilities, MC is a suitable candidate for comparison.

MC relies on assumptions 1 and 2, as introduced in the main document. This is necessary, since it is imperative to compare both BP and MC such that they are tasked to solve the same problem with the same likelihood function over an identically sized search space. On the other hand, neither assumptions 3 nor 4 are not applicable to the MC algorithm, as MC is a search over discrete parameter configurations. Therefore, comparing BP results against MC results might isolate the effects of assumptions 1 and 2, collectively, on the quality. It should be noted that letting MC run for longer times would improve MC performance.

We encode a termination criterion that stops when the algorithm has found a strongly attracting minimum for a pre-fixed number of iterations. Our MC algorithm keeps track of the top 1,000 configurations it has encountered, as evaluated by the log-likelihood (or cost function) in Equation 6b. MC traverses solution space by drawing each parameter from a zero-heavy distribution, where all non-zero weights are given equal probability. When the current configuration has lower cost than the previous configuration, it is accepted and appended to the list of the top 1000 configurations, while the highest cost configuration in the list of the top 1000 is expelled. In order to prevent the search from getting stuck in a weakly attracting local-minimum, we consider accepting higher cost configurations with probability:

$$P(keep\ Wnew) = \frac{e^{C(W_{new})}}{e^{C(W_{current})}}$$

The algorithm terminates when the list of the top 1000 configurations does not change after 1000 iterations.

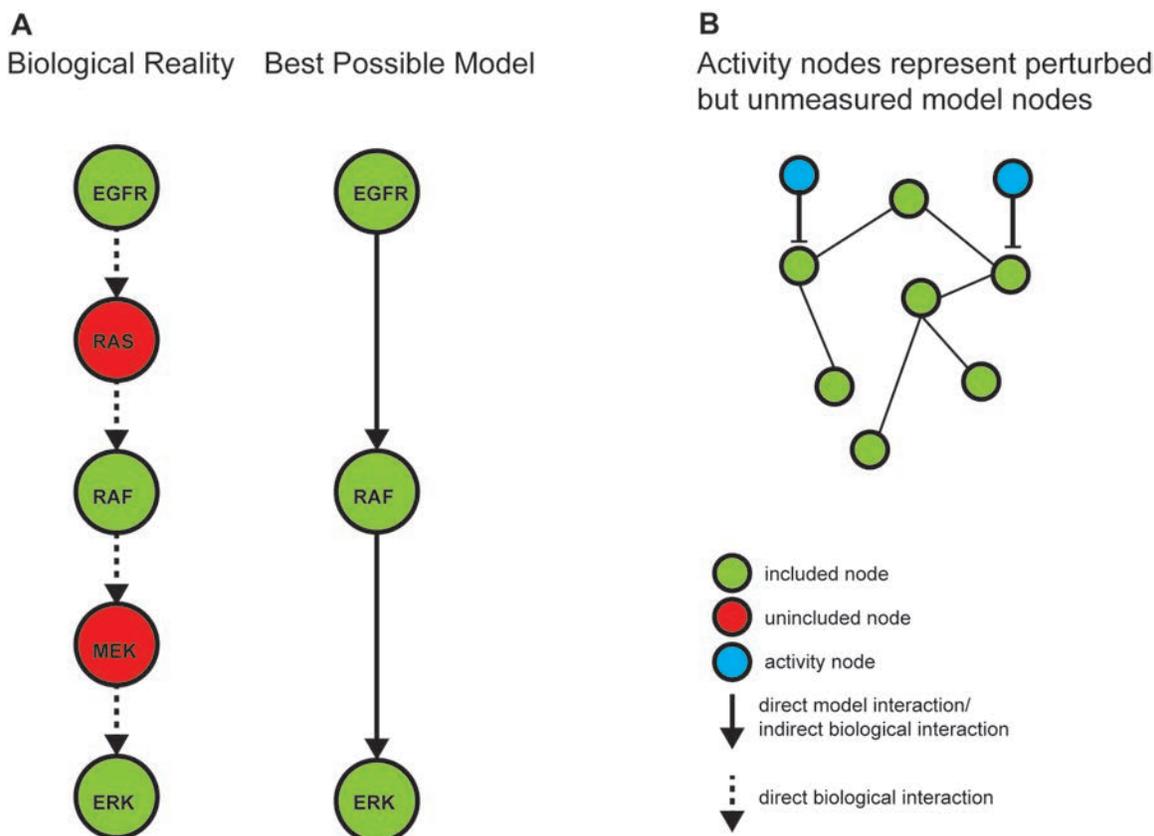

**Figure S-15: Schematic illustration of interpreting model edges and activity nodes.** Direct interactions in our models do not necessarily imply direct biological interactions (A). Rather the separation between any two connected nodes is dependent on those nodes that are included in the model, which is in turn a function of the availability and specificity of assays for various proteins and phospho-proteins. Consider the linear cascade of 5 nodes (A). When intermediate nodes are excluded from the model, for whatever reason, the true direct interactions are also excluded. In their stead, the model will have interactions between the neighboring nodes. In that case, the direct model interactions do no correspond to direct biological interactions. The use of activity nodes may also confuse the interpretation of our model interactions. Activity nodes stand in for the activity of a protein (or phosphorylated protein) that is not directly measureable, as explained in the Materials and Methods section of the main manuscript. It is this activity that is being targeted by a given drug. We can assume that the activity is below basal (x is negative) when the drug is applied. However, in those conditions in which the drug is not applied we have no data or reasonable assumption with which to approximate the activity nodes. Consequently we do not have enough data with which to infer interactions into activity nodes. Activity nodes are therefore restricted to exist as 'root nodes' such that they have only outgoing edges (B). We consider the activity nodes to represent the effect of ad rug on the rest of the model nodes.

**Supplementary Text S6: BP Implementation considerations.**

The main manuscript covers all of the theoretical considerations of our BP algorithm. However, there are a few implementation issues that affect the efficiency of the algorithm.

Computational calculation of Equation 14.A for each update requires substantial memory and computational power. The algorithm has to keep track of all M x(N -1) factors, and then perform *M* multiplications, (*N*-1) times in preparation for a single update. We take advantage of the multiplicative relationship between a marginal P(wij ) and its factors $\rho^\mu(w_{ij})$, to reduce the number of computations need to evaluate $P^\mu(w_{ij})$ and eliminate precision error associated with many multiplications of small numbers.

The marginal distribution and the global information are related by a simple division of the marginal by the cavity factor (Equations S1A and S1B). In log space, the same calculation is just a simple subtraction (Equation S1C).

**Equation S1**

$$P(w_{ij}) = e^{-\lambda\delta(w_{ij})} \prod_v^M \rho^v(w_{ij})$$

$$P^\mu(w_{ij}) = \frac{P(w_{ij})}{\rho^\mu(w_{ij})}$$

$$\ln P^\mu(w_{ij}) = \ln P(w_{ij}) - \ln \rho^\mu(w_{ik})$$

$$= -\lambda\delta(w_{ij}) + \sum_v^M \ln \rho^v(w_{ij}) - \ln \rho^\mu(w_{ik})$$

The calculation in equation S1C is a fast calculation and involves only an inverse log operation, one subtraction and the cost of doing the normalization (K additions and K division operations). The calculation needs only to keep track of all factors and the sum of the natural logarithms of the factors across all M conditions, stored in an array $\Psi$.

**Equation S2**

$$\Psi(j) = \sum_v^M \ln\left(\rho^v(w_{ij})\right) - \lambda\delta(w_{ij})$$

$$P^\mu(w_{ij}) = \frac{1}{Z}\left(e^{\Psi(j) - \ln(\rho^\mu(w_{ik}))}\right)$$

An additional advantage for performing the calculation in log space is that the operation does not risk precision errors from the multiplication of many small numbers. The calculation is a subtraction of negative numbers followed by a natural antilogarithm calculation, neither of which risk precision error. In non log space, the calculation requires M multiplications of probability distributions, which contain some very small numbers. In fact, an earlier implementation of the BP algorithm experienced detrimental errors for even small M. This error cascaded and produced meaningless results. The log space calculation is robust up to about at least M=1,000, above which has not been tested.

**Analytical approximation of the local factor update**

The integral in Equation 14B is not solvable analytically due to the nonlinearity introduced from the sigmoid function $\phi(z)$. One can solve the integral numerically with a method such as trapezoidal integration. However, since Equation 14B has to be calculated NxM times per iteration over many iterations, numerical solutions of this kind dramatically slow down the computational time to convergence. To alleviate the cost of this calculation, we use a simple algebraic rearrangement as shown in Equation S4B to make the fitness term $F^\mu(s_k^\mu)$ linear with the integration variable $s_k^\mu$.

The goal is to rewrite the error inside the sum of squares term to be linear in $s_k^\mu$. The fitness is maximum when the cost is zero and therefore $x_i^{*\mu} = x_i^\mu$, which in turns means that $x_i^{*\mu} = \phi(s_k^\mu + w_{ik} x_k^{*\mu})$. Inverting the nonlinear function on both sides of the equation (Equation S3A) and yields an approximation of the fitness that is linear with $s_k^\mu$.

**Equation S3**

$$\phi^{-1}(x_i^{*\mu}) - w_{ik} x_k^{*\mu} = s_k^\mu$$

$$F^\mu(s_k^\mu) = e^{([\phi^{-1}(x_i^{*\mu}) - w_{ik} x_k^{*\mu}] - s_k^\mu)^2}$$

Equation S4 is the analytical solution to the integral in Equation S3B.

**Equation S4**

$$\rho^\mu(w_{ik}) = e^{\frac{-\beta w_{ik} x_k^{*\mu} + [-\phi^{-1}(x_i^{*\mu}) - \overline{s_k^\mu}]}{2\beta\sigma^2 + 1}}$$

**The Algorithm**
See Figure-S16 for an algorithm flow-chart.

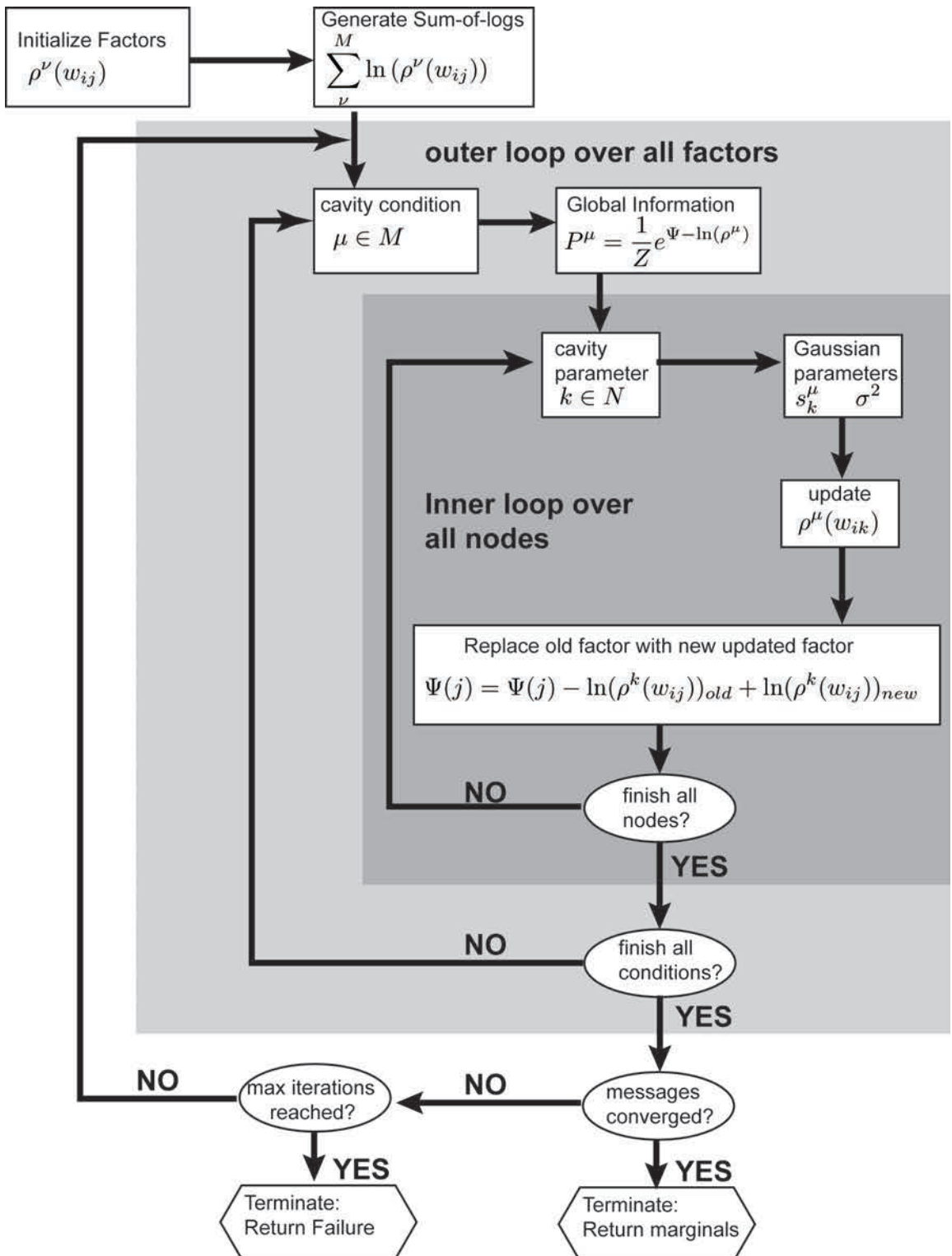

**Figure S-16: Algorithm flow chart.**

**Supplementary Table S1: Quantitative details and biological significance of inferred interactions**

| Modeled Interaction | Confirmed | <Wij> | Notes |
|---|---|---|---|
| 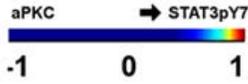 aPKC → STAT3pY705 | Y | 1.00 | PKC mediates the phosphorylation of STAT3 at Y727 following STAT3 phosphorylation at Y705 and dimerization. Thus, this is a potential logical interaction. |
| 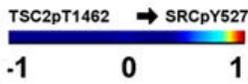 TSC2pT1462 → SRCpY527 | N | 0.97 | A predicted interaction. Possible negative feedback acting from PI3K/AKT pathway to upstream SRC. |
| 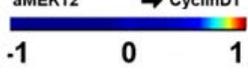 aMEK12 → CyclinD1 | Y | 0.94 | Indirect/logical interaction. RAF/MEK/ERK signaling induces Cyclin D1 expression and regulates the transition from G1 to S phase. |
| 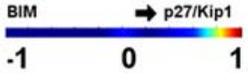 BIM → p27/Kip1 | N | 0.92 | P27/Kip1 is a tumor suppressor that inhibits cyclin D1. BIM is a proapoptotic protein. The response profiles of the two proteins are highly correlated. The inferred interaction reflects this correlation. Observed edge may be due to a logical interaction. Both proteins are co-regulated by FoxO transcription factor in response to IL-2. |
| 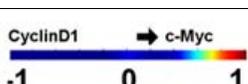 CyclinD1 → c-Myc | Y | 0.89 | The inferred edge between CyclinD1 and c-Myc is bidirectional. c-Myc transcription factor regulates expression of a large spectrum of oncogenic proteins including cyclins. The response profiles of CyclinD1 and c-Myc are highly correlated. However, the inverse regulation is not reported in literature. |
| 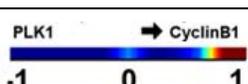 PLK1 → CyclinB1 | Y | 0.87 | PLK1 and CyclinB1 interaction is critical for G2/M transition in cell cycle. This is validated as a direct interaction. |
| 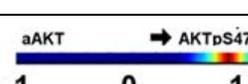 aAKT → AKTpS473 | Y | 0.86 | Element of PI3K/AKT pathway. aAKT is an activity node, which spesifically corresponds to localization of AKT to membrane through PIP3 binding in our context. After membrane recruiutment, phosphorylation at S473 activates AKT. |
| 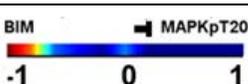 BIM ⊣ MAPKpT202 | Y | -0.85 | BIM and MAPK bidirectional interaction is inferred in network models. Multiple studies suggest that proapoptotic activity of BIM is regulated by MAPK through phosphorylation at multiple sites. |
| 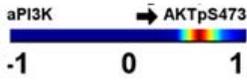 aPI3K → AKTpS473 | Y | 0.69 | Well established logical interaction. PI3K facilitates the membrane localization and phosphorylation of through PIP3 formation. Once localized, AKT is phosphorylated at T308 by PI3K dependent kinase 1 (PDK1). This phosphorylation is followed by a second phosphorylation at S473 by mTOR-rictor complex. |
| 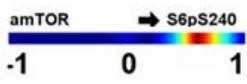 amTOR → S6pS240 | Y | 0.68 | Part of PI3K/AKT pathway. amTOR is an activity node. Phosphorylation at S240 activates S6. This is an indirect interaction that takes place through p70S6K (not included in the model.). S56 activation triggers Protein translation. |
| 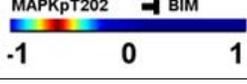 MAPKpT202 ⊣ BIM | Y | -0.66 | MAPK phosphorylates BIM on multiple Serine sites leading to degradation of BIM. |
| 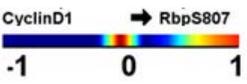 CyclinD1 → RbpS807 | Y | 0.62 | Phosphorylation of tumor suppressor protein Rb at S807 by CDK4/6 leads to its inhibition of Rb and G1/S transition. CDK4/6 activation by Cyclin D1 has a major role in inducing Rb phosphorylation. |
| 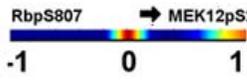 RbpS807 → MEK12pS217 | N | 0.59 | In canonical RAF/MEK/ERK pathway, an indirect, logical interaction could be defined from MEK to Rb deactivation. However, this interaction has an opposite direction. Few studies point an indirect, bidirectional genetic interaction with Rb and N-RAS, which is upstream of MEK. The observed edge could be a false positive due to the high experimental correlation observed in response profiles of two proteins or may reflect a highly complicated, indirect interaction between MEK and Rb. |
| 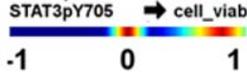 STAT3pY705 → cell_viabil | Y | 0.57 | PKC regulates STAT3 phosphorylation (see above) and PKC inhibition leads to cell death. The inferred edge reflects the influence of PKC inhibition on cell viability through STAT3 activity. |
| 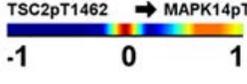 TSC2pT1462 → MAPK14pT180 | N | 0.56 | The reverse interaction is reported as an indirect event. This is a novel prediction and requires experimental validation. |
| 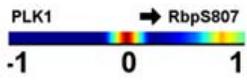 PLK1 → RbpS807 | Y | 0.54 | PLK1/CyclinB interactions function to activate CDK1 activation, which in turn phosphorylates and activates tumor suppressor Rb. An inverse interaction is reported such that Rb activation leads to the |

| | | | suppression of PLK1 expression. |
|---|---|---|---|
| 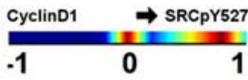 | N | 0.54 | This is a possible negative feedback loop acting on Src from Cyclin D1. Src is deactivated when phosphorylated at Y527. Src is upstream of multiple pathways, whose activation lead to increase in Cyclin D1 expression. |
| 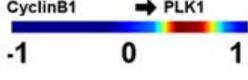 | Y | 0.52 | CyclinB1-CDK1 and PLK1 activity induce G2/M transition in cell cycle. PLK1 is an important target for anticancer drugs. CyclinB-CDK1 complex and PLK1 regulate the activity of each other. However, the activation of PLK1 by CyclinB1 is most probably an indirect process in mammals and the precise functional relationship between them is highly dependent on biological context. Note that in X. leavis CDK1 directly phosphorylates PLK1 At S340 (not conserved in mammals) and induces its activation. |
| 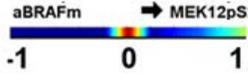 | Y | 0.50 | Part of canonical BRAF/MEK/ERK pathway. |